\documentclass[11pt,a4paper]{article}
\pdfoutput=1

\usepackage[margin=1in]{geometry}

\let\bibhang\relax
\usepackage[authoryear]{natbib}
\usepackage[T1]{fontenc}
\usepackage[utf8]{inputenc}
\usepackage{pmboxdraw}
\usepackage{graphicx}
\usepackage{subcaption}
\usepackage{fancyvrb}
\usepackage{amsmath, mathtools, amssymb, bm}

\newcommand{\x}{\boldsymbol{x}}
\newcommand{\z}{\boldsymbol{z}}

\renewcommand{\b}{\boldsymbol{b}}
\newcommand{\Deltab}{\boldsymbol{\Delta}}

\newcommand{\A}{\boldsymbol{A}}
\newcommand{\U}{\boldsymbol{U}}

\newcommand{\mub}{\boldsymbol{\mu}}
\newcommand{\Sigmab}{\boldsymbol{\Sigma}}

\newcommand{\lambdab}{\boldsymbol{\lambda}}
\newcommand{\T}{^{\top}}

\newcommand{\iset}{\mathcal{I}}
\newcommand{\Real}{\mathbb{R}}
\DeclareMathOperator*{\argmax}{arg\max}

\makeatletter
\newcommand\code{\bgroup\@makeother\_\@makeother\~\@makeother\$\@codex}
\def\@codex#1{{\normalfont\ttfamily\hyphenchar\font=-1 #1}\egroup}
\makeatother

\begin{document}

\title{A fast and efficient Modal EM\\
       algorithm for Gaussian mixtures -- pippo}
\author{Luca Scrucca\\
        Dipartimento di Economia, Università degli Studi di Perugia, Italy}
\date{\today}

\maketitle

\begin{abstract}
In the modal approach to clustering, clusters are defined as the local maxima of the underlying probability density function, where the latter can be estimated either non-parametrically or using finite mixture models. Thus, clusters are closely related to certain regions around the density modes, and every cluster corresponds to a bump of the density.
The Modal EM algorithm is an iterative procedure that can identify the local maxima of any density function. In this contribution, we propose a fast and efficient Modal EM algorithm to be used when the density function is estimated through a finite mixture of Gaussian distributions with parsimonious component-covariance structures. After describing the procedure, we apply the proposed Modal EM algorithm on both simulated and real data examples, showing its high flexibility in several contexts.\\

\end{abstract}

\newpage
\baselineskip=18pt

\section{Introduction}
\label{sec:intro}

The term cluster analysis encompasses a large set of methods and algorithms that aim at partitioning a set of data into some meaningful groups of homogeneous data points called \emph{clusters}. The presence of such clusters is not known a priori, sometimes even their number is unknown, nor is case labelling available. For this reason, cluster analysis is considered an instance of so-called \emph{unsupervised learning}. 

Several approaches and methods are available in the literature to explore the clustering structure of a dataset \citep{Everitt:etal:2011}. Among these, density-based approaches have been proposed to exploit the relationship between the underlying density of a dataset and the presence of clusters. 
In the parametric or \emph{model-based clustering} approach each component of a mixture distribution is associated to a cluster \citep{McLachlan:Peel:2000, Fraley:Raftery:2002}. Thus, observations are allocated to the cluster with maximal weighted component density. 

However, there may be situations where more than a single component is required to represent the shape of a cluster. Merging of mixture components is a possible answer to this problem. \citet{Baudry:etal:2010} proposed a merging method based on an entropy criterion, while \citet{Hennig:2010} discussed several methods based on unimodality and misclassification probabilities. All these methods are hierarchical in nature, so clusters can only be obtained by merging two or more mixture components. This indeed may constitute a limitation because data points assigned to a single Gaussian component cannot be subsequently allocated to different clusters.
A different approach to tackle this problem was proposed by \citet{Scrucca:2016} based on the identification of connected components from high density regions of the underlying density function.

\emph{Modal clustering} is another density-based approach to clustering where clusters are taken as the ``domains of attraction'' of the density modes \citep{Stuetzle:2003}.   
This follows the definition proposed by \citet[][p. 205]{Hartigan:1975}, according to which ``clusters may be thought of as regions of high density separated from other such regions by regions of low density''. This definition of cluster is the one adopted in the paper.

Modal EM (MEM) is an iterative algorithm aimed at identifying the local maxima of a density function \citep{Li:Ray:Lindsay:2007}. 
Let $f(\x) = \sum_{k=1}^G \pi_k f_k(\x)$ be a finite mixture density for $\x \in \Real^d$, where $\pi_k$ is the mixing probability of component $k$ with density function $f_k(\x)$, under the constraints $\pi_k > 0$ for all $k=1,\ldots,G$, and $\sum_{k=1}^G \pi_k = 1$. 
Given an initial starting point $\x^{(0)}$, the following steps are iteratively executed until a stopping criterion is met:
\begin{eqnarray*}
\text{E-step:} & 
p_k^{(t)} & = \frac{\pi_k f_k(\x^{(t-1)})}{f(\x^{(t-1)})}
\qquad \text{for } k = 1, \ldots, G; \\
\text{M-step:} & 
\x^{(t)} & = \argmax_{\x} \sum_{k=1}^G p_k^{(t)} \log f_k(\x^{(t-1)}).
\end{eqnarray*}

\citet{Li:Ray:Lindsay:2007} showed that the objective function in the M-step has a unique maximum if the $f_k(\x)$ are Gaussian densities. 
They also reported a closed-form solution in case of Gaussian mixtures with common covariance matrix. This is a fairly strong assumption that rarely occurs in practice, so it would be interesting to address the general case, which is not only more complex to deal with, but also much more interesting from a practical point of view.   

Regarding the question of how many modes a Gaussian mixture can have, we note that \citet{Carreira-Perpinan:Williams:2003a, Carreira-Perpinan:Williams:2003b} conjectured that the number of modes cannot exceed the number of components when the components of the mixture have the same covariance matrix (homoscedastic mixture), whereas if the components are allowed to have arbitrary and different covariance matrices (isotropic and full heteroscedastic mixtures) then the number of modes can be larger than the number of components. 
However, the first conjecture turns out to be wrong, so in general it is not possible to know a priori the number of modes of a multivariate Gaussian mixture. For a recent contribution on this issue see \citet{Amendola:etal:2020}, where lower and upper bounds on the maximum number of modes of a Gaussian mixture are derived under the assumption they are finite.

Modal clustering plays a central role in the non-parametric approach to cluster analysis. 
Several mode-seeking algorithms have been proposed in the literature, such as the mean-shift algorithm of \citet{Fukunaga:Hostetler:1975} and its many extensions \citep{Carreira-Perpinan:2016}.
However, regardless of the algorithm adopted, detection of high-density regions requires the choice of a density estimator, typically a kernel density estimator. 
The latter requires the selection of an appropriate kernel bandwidth, and extension to high dimensions is known to be somewhat problematic \citep[ch. 9]{Scott:2009}.
Interestingly, connections exist between the Modal EM algorithm and the mean shift algorithm. In fact, \citet{Carreira-Perpinan:2007} showed that the mean-shift algorithm is a generalized EM algorithm when the kernel of a non-parametric kernel density estimate is Gaussian. More recently, \citet{Chacon:2019} extended the use of the mean shift algorithm to non-isotropic Gaussian components. 
For a review on non-parametric modal clustering see \citet{Menardi:2016}.

\begin{figure}[htb]
\centering
\subcaptionbox{}{\includegraphics[width=0.33\textwidth]{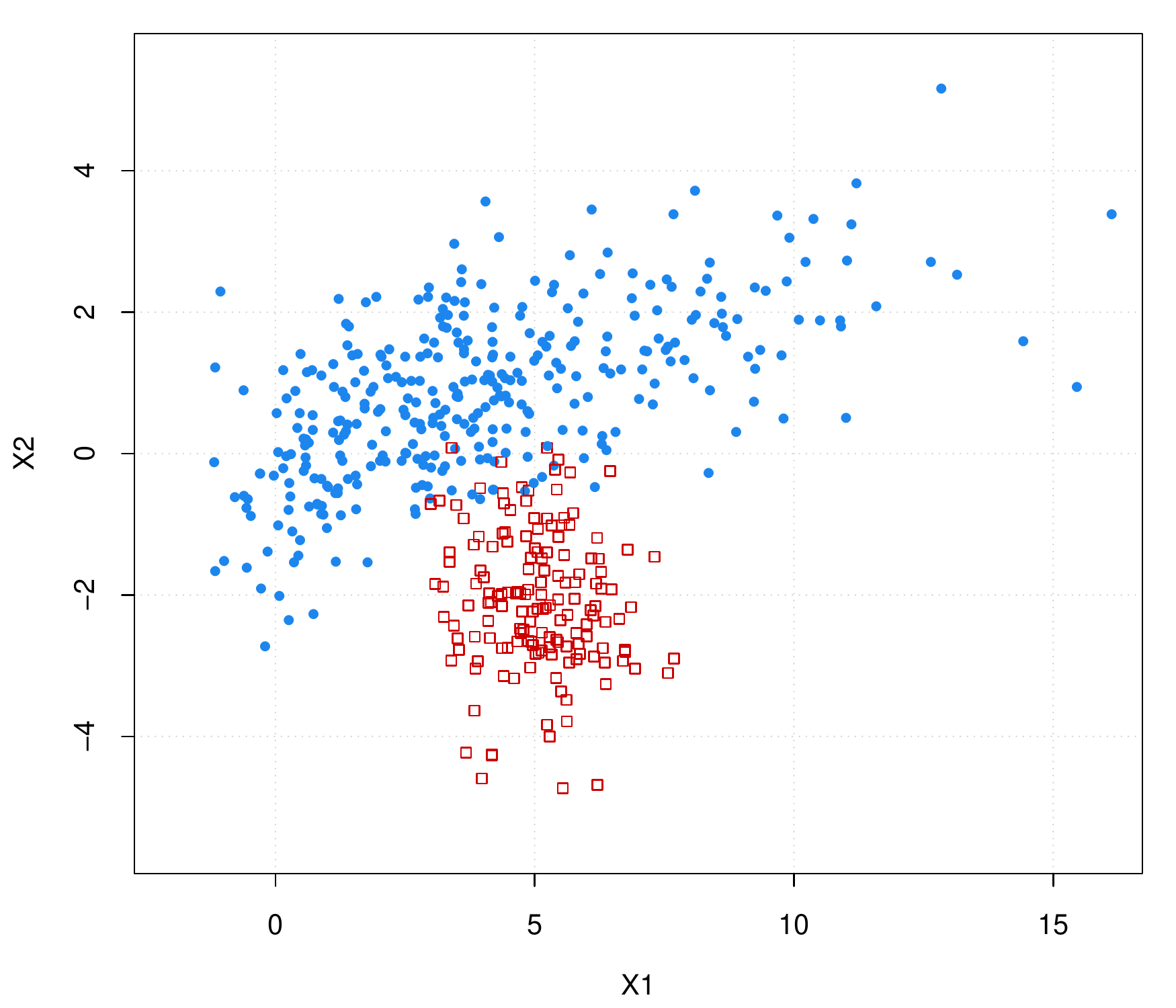}}%
\subcaptionbox{}{\includegraphics[width=0.33\textwidth]{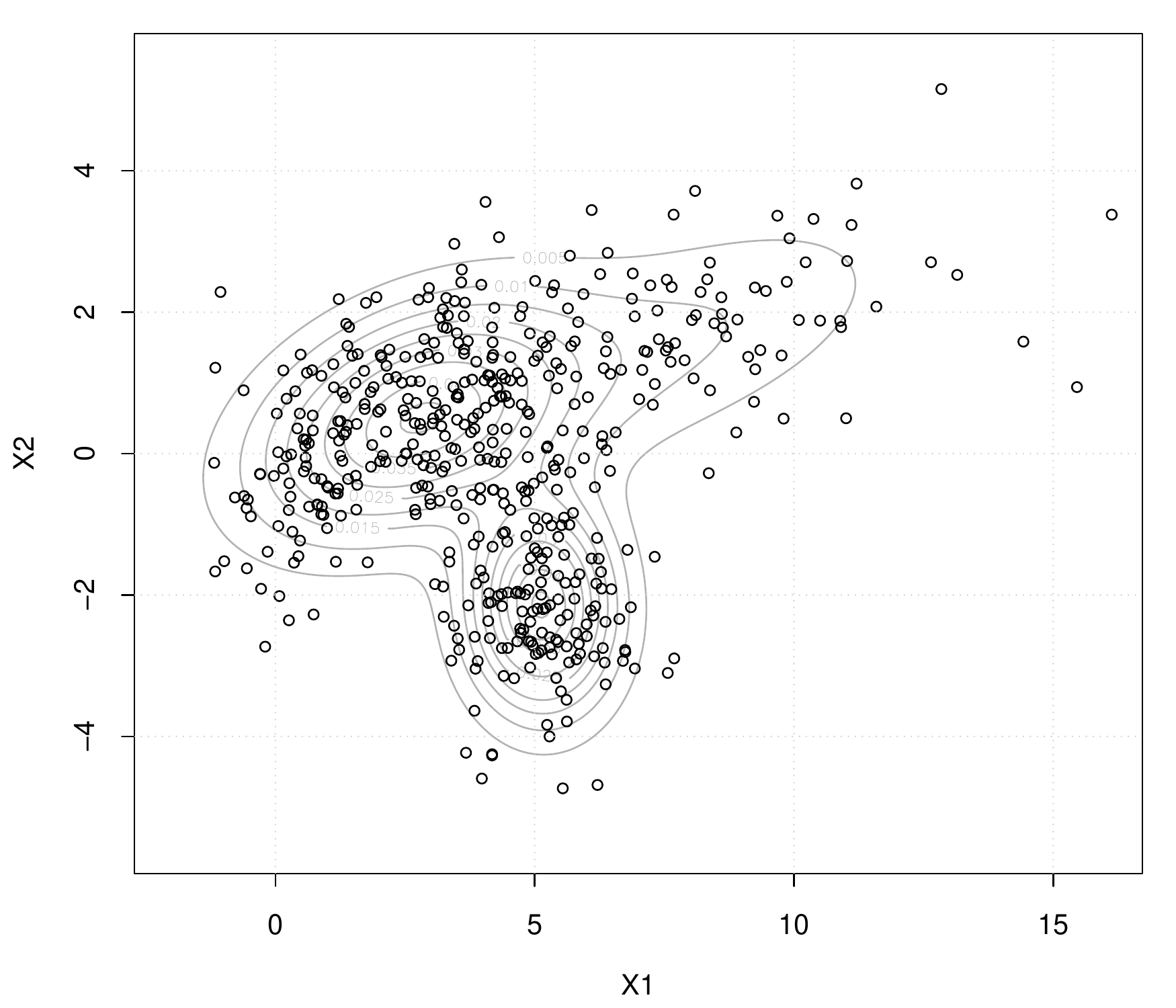}}%
\subcaptionbox{}{\includegraphics[width=0.33\textwidth]{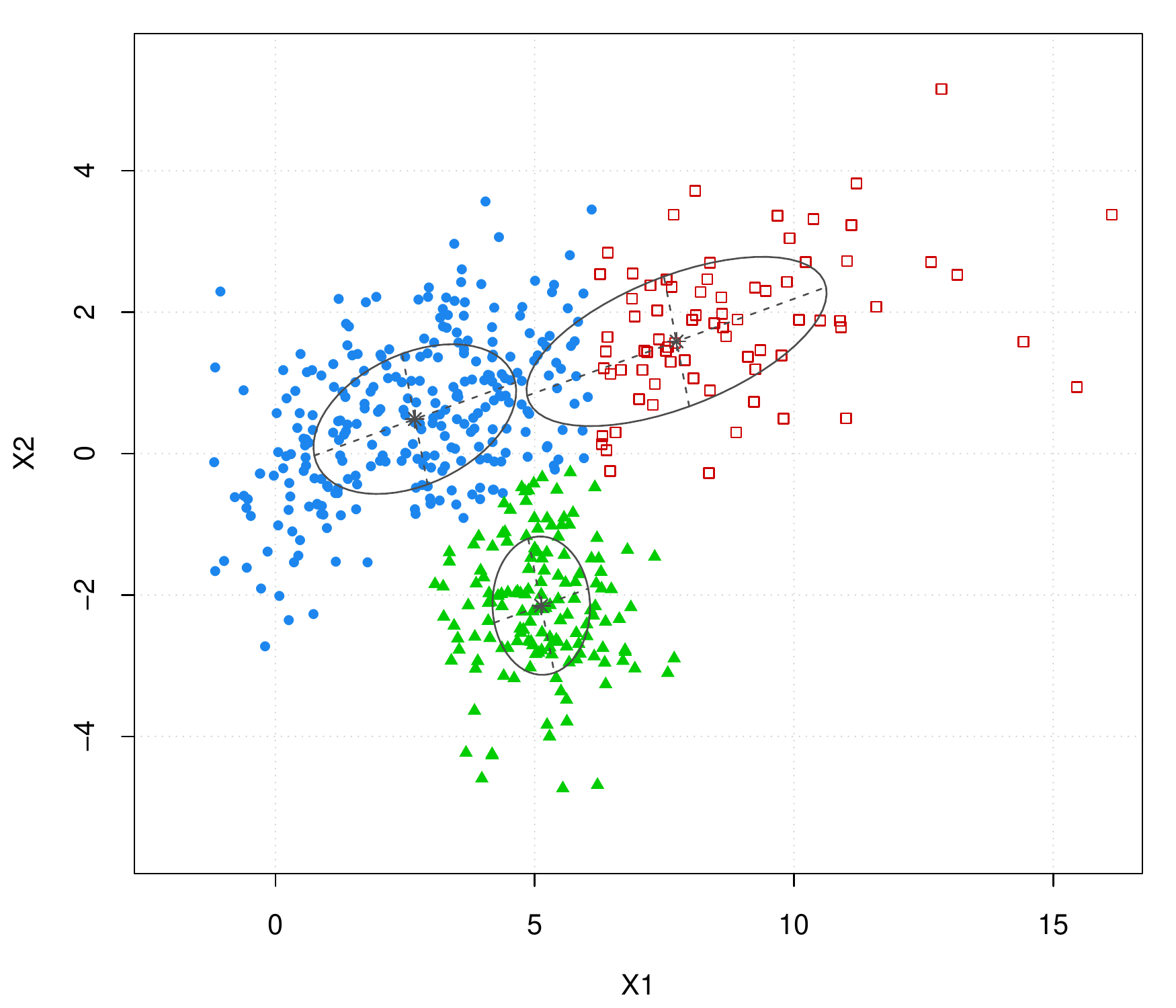}}%
\caption{Plots of a two-component simulated data: (a) data points marked according to the true component memberships; (b) density contours obtained from the estimated Gaussian mixture model; (c) clustering and ellipses corresponding to the estimated Gaussian components of the mixture.}
\label{fig1:motiv_example}
\end{figure}

\paragraph{A motivating example} 
Consider the data shown in Figure~\ref{fig1:motiv_example}a. They represent a sample of $n = 500$ observations drawn from the following bivariate two-component mixture:
$$
f(\x) = \pi \; N(\mub_1, \Sigmab_1) + (1-\pi) \; \text{Skew}N(\mub_2, \Sigmab_2, \lambda_2),
$$
where $\pi = 1/3$ is the mixing weight of the first Gaussian component with mean $\mub_1 = [5\; -2]\T$ and covariance matrix $\Sigmab_1 = \left[\begin{smallmatrix} 1 & 0 \\ 0 & 1 \end{smallmatrix}\right]$,
whereas the second component is a Skew-Normal distribution \citep{Azzalini:2013} with location $\mub_2 = [0 \; 0]\T$, scale matrix $\Sigmab_2 = \left[\begin{smallmatrix} 1 & 0.5 \\ 0.5 & 1 \end{smallmatrix}\right]$, and skew parameter $\lambdab_2 = [5 \; 1]\T$.
Figure~\ref{fig1:motiv_example}b shows the density estimate corresponding to the ``best'' Gaussian finite mixture model according to BIC. The selected model is a mixture of three components with ellipsoidal covariance matrices having common orientation (VVE in \texttt{mclust} nomenclature; see \citet{Scrucca:etal:2016}).
Figure~\ref{fig1:motiv_example}c shows the corresponding clustering partition. 
Clearly, observations coming from the skewed component are not correctly identified by the estimated clustering partition. Indeed, two Gaussian components are needed to adequately represent this group of observations. However, note that the corresponding density estimate seems to correctly suggest a bimodal distribution. By exploiting this fact a better partition could be obtained, and the method discussed in this paper aims to deal with similar situations.

\medskip

In this contribution we propose a fast and efficient Modal EM algorithm for identifying the modes of a density estimated by finite mixture of multivariate Gaussians having any of the parsimonious covariance structures available in the \texttt{mclust} R package \citep{Scrucca:etal:2016}.
The outline of this article is as follows.
Section 2 provides a brief review of the Modal EM approach for Gaussian mixtures available in the literature. Section 3 contains the proposal for extending the 
Modal clustering approach to any density estimated by fitting a finite mixture of Gaussian distributions with parsimonious component-covariance structures, and details on how to improve the computational efficiency of this approach. 
Section 4 describes the empirical results deriving from the application of the proposed Modal EM algorithm to examples using both synthetic and real datasets.
The final section provides some concluding remarks.

\section{Modal EM algorithm for Gaussian mixtures}
\label{sec:MEMGMM}

Gaussian mixture models (GMMs) assume that the mixture components are all multivariate Gaussians with mean $\mub_k$ and covariance $\Sigmab_k$, i.e. $f_k(\x) \equiv \phi(\x ; \mub_k, \Sigmab_k)$. 
Therefore, the mixture density for any data point $\x_i$ can be written as
\begin{equation*}
f(\x_i) = \sum_{k=1}^G \pi_k \phi(\x_i ; \mub_k, \Sigmab_k).
\end{equation*}
Clusters described by a GMM are centred at the means $\mub_k$, and with other geometric characteristics (such as volume, shape and orientation) determined by the covariance matrices $\Sigmab_k$. These can be controlled by introducing some constraints on the covariance matrices through the following eigen-decomposition \citep{Banfield:Raftery:1993, Celeux:Govaert:1995}
\begin{equation}
\Sigmab_k = \lambda_k \U_k \Deltab_k \U\T_k,
\label{eq:eigendecomp}
\end{equation}
where 
$\lambda_k = |\Sigmab_k|^{1/d}$ is a scalar which controls the \emph{volume},
$\Deltab_k$ is a diagonal matrix, such that $|\Deltab_k| = 1$ and with the normalised eigenvalues of $\Sigmab_k$ in decreasing order, which controls the \emph{shape},
$\U_k$ is an orthogonal matrix of eigenvectors of $\Sigmab_k$ which controls the \emph{orientation}. In this way, a total of 14 GMMs are obtained \citep{Scrucca:etal:2016}.

It is important to note that in this paper we shall consider the mixing proportions $\pi_k$, the mean vectors $\mub_k$, and the covariance matrices $\Sigmab_k$ as fixed (either estimated or known a priori) for all $k=1, \ldots, G$. 

The MEM algorithm starts with $t = 0$ and initial data point $\x_i^{(0)} = \x_i$. At iteration $t$, MEM performs the following steps:
\begin{itemize}

\item Set $t = t + 1$.

\item E-step -- update the posterior conditional probability of the current data point $\x_i$ to belong to the $k$th mixture component:
\begin{equation*}
z_{ik}^{(t)} = \frac{\pi_k \phi(\x_i^{(t-1)} ; \mub_k, \Sigmab_k)}{ \sum_{g=1}^G \pi_g \phi(\x_i^{(t-1)} ; \mub_g, \Sigmab_g) },
\end{equation*}
for all $k = 1, \ldots, G$.

\item M-step -- update the current value of $\x_i$ by solving the optimisation problem:
\begin{equation*}
\x_i^{(t)} = \argmax_{\x_i} \sum_{k=1}^G z_{ik}^{(t)} \log \phi(\x_i^{(t-1)} ; \mub_k, \Sigmab_k).
\end{equation*}

\item Iterate the above steps until a stopping criterion is satisfied, for instance 
$\max \{ |\x_i^{(t)} - \x_i^{(t-1)}| / (1 + |\x_i^{(t-1)}|) \} < \epsilon$, where $\epsilon$ is a tolerance value, say $\epsilon = 1\text{e-}5$,
or a pre-specified maximum number of iterations is reached.
\end{itemize}

By the ascending property of the MEM algorithm \citep[][Appendix A]{Li:Ray:Lindsay:2007}, at convergence the value $\x_i^{(t)}$ is the mode associated with data point $\x_i$. 
\citet{Li:Ray:Lindsay:2007} presented a closed-form solution only in the specific case of Gaussian mixtures with common covariance matrix, and reported that numerical procedures are required for the M-step if the covariance matrices are different across components. 
By replicating the above algorithm for all data points, it is possible to identify the modes associated with any $\x_i$ ($i = 1, \ldots, n$), but this process is time-consuming even for moderately large datasets. 
In the next section we present an approach aimed at accelerating the MEM algorithm by iterating simultaneously for all data points and for any parsimonious covariance matrix decomposition. 

\section{Proposal}
\label{sec:MEMGMMfast}

In this section we detail our proposal to speed up the MEM algorithm for Gaussian mixtures having any of the parsimonious component-covariance matrix eigen-decomposition proposed by \citet{Banfield:Raftery:1993, Celeux:Govaert:1995}, and implemented in the \texttt{mclust} package \citep{Scrucca:etal:2016} for R \citep{Rstat}.

To this end, we start by noting that, the objective function in the M-step presented in Section~\ref{sec:MEMGMM} can be written as 
\begin{equation*}
Q(\x_i) = \sum_{k=1}^G z_{ik} \log \phi(\x_i ; \mub_k, \Sigmab_k).
\end{equation*}
The gradient and Hessian of this function with respect to the observed vector $\x_i$ (again, assuming the mixture parameters $\{\pi_k, \mub_k, \Sigmab_k\}_{k=1}^G$ as known and fixed) are, respectively,
\begin{align*}
\nabla Q(\x_i) & = - \sum_{k=1}^G z_{ik} \Sigmab_k^{-1} (\x_i - \mub_k), \\
\shortintertext{and}
\nabla^2 Q(\x_i) & = - \sum_{k=1}^G z_{ik} \Sigmab_k^{-1}.
\end{align*}
Because all covariance matrices $\Sigmab_k$ are positive definite by definition, and $z_{ik} > 0$ for all $k$ and $i$, the Hessian is negative definite.
Thus, maximisation of the $Q$-function can be pursued by equating the gradient to zero, and then solving for $\x_i$ we obtain
\begin{equation}
\x_i^* = \left( \sum_{k=1}^G z_{ik} \Sigmab_k^{-1} \right)^{-1} \sum_{k=1}^G z_{ik} \Sigmab_k^{-1} \mub_k.
\label{eq:mem_mstep_opt}
\end{equation}
Note that the last equation also arises in other mode-seeking procedures, such as in the gradient-quadratic algorithm and fixed-point iterative algorithm proposed by \citet{Carreira-Perpinan:2000}, and the mean shift algorithm proposed by \citet{Chacon:2019}. 

A straightforward application of equation~\eqref{eq:mem_mstep_opt} requires to replicate the procedure for all data points. This can be time-consuming because it repeatedly involves calculating matrix products and inversion of matrices.
However, these objects can be efficiently computed in a single pass for all data points through the use of the Kronecker product.

Let $\z_k$ be the vector of length $n$ containing the posterior probabilities of all data points $\{ \x_i \}_{i=1}^n$ to belong to the $k$th mixture component, and $\mub_k$ be the vector of length $d$ of component means ($k = 1, \ldots, G$). 
Define the $(nd \times d)$ matrix 
$$
\A = \sum_{k=1}^G \z_k \otimes \Sigmab_k^{-1},
$$
and the $(nd \times 1)$ vector 
$$
\b = \sum_{k=1}^G \z_k \otimes \Sigmab_k^{-1} \mub_k.
$$ 
Solutions for each $\{ \x_i \}_{i=1}^n$ can be obtained by solving the linear systems 
$$
\A_{\iset} \x_i^* = \b_{\iset},
$$
where $\iset \equiv \{(i-1)d+1, \ldots, id \}$ is the set containing the indices used to select the rows of matrix $\A$ and the elements of vector $\b$. Equivalently, solutions of the linear systems can be written as $\x_i^* = \A_{\iset}^{-1} \b_{\iset}$.

Compared to the approach based on the calculation of the solution for each data point as in \eqref{eq:mem_mstep_opt}, the main advantage of our proposal is that computing the matrix $\A$ and vector $\b$ is performed in a single step for all the observations. Then, the use of the indices $I$ allows us to select the relevant parts of $\A$ and $\b$ for computing the solutions. Although algebraically equivalent, this approach turns out to be three times faster in our experiments under different settings. 

The above algorithm is fast and efficient, but in practice Modal EM can suffer from some drawbacks which can be easily addressed as discussed below.

\subsection{Setting the step size}
\label{sec:MEMGMM_stesize}

\begin{figure}[htb]
\centering
\subcaptionbox{}{\includegraphics[width=0.45\textwidth]{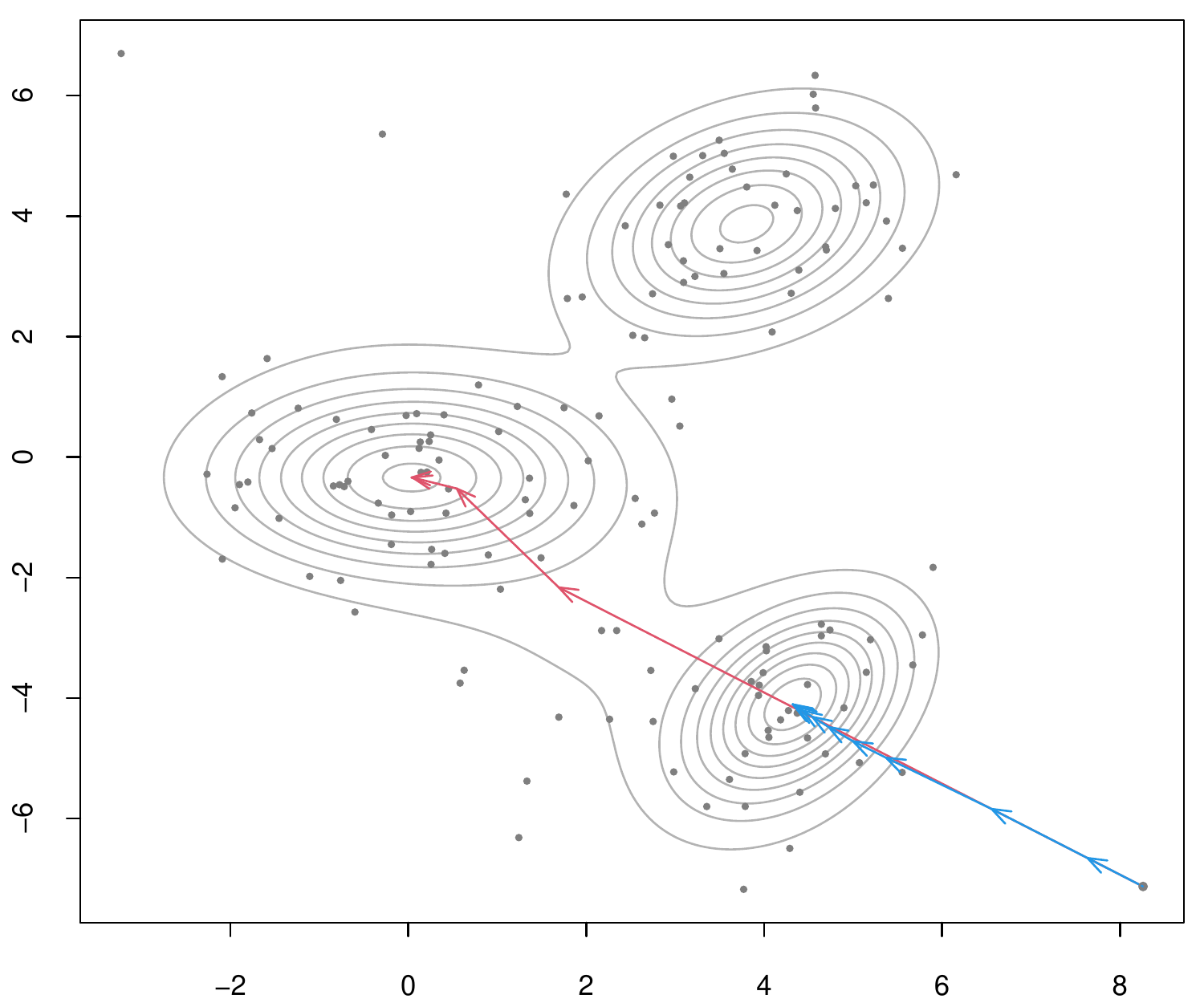}}\;
\subcaptionbox{}{\includegraphics[width=0.45\textwidth]{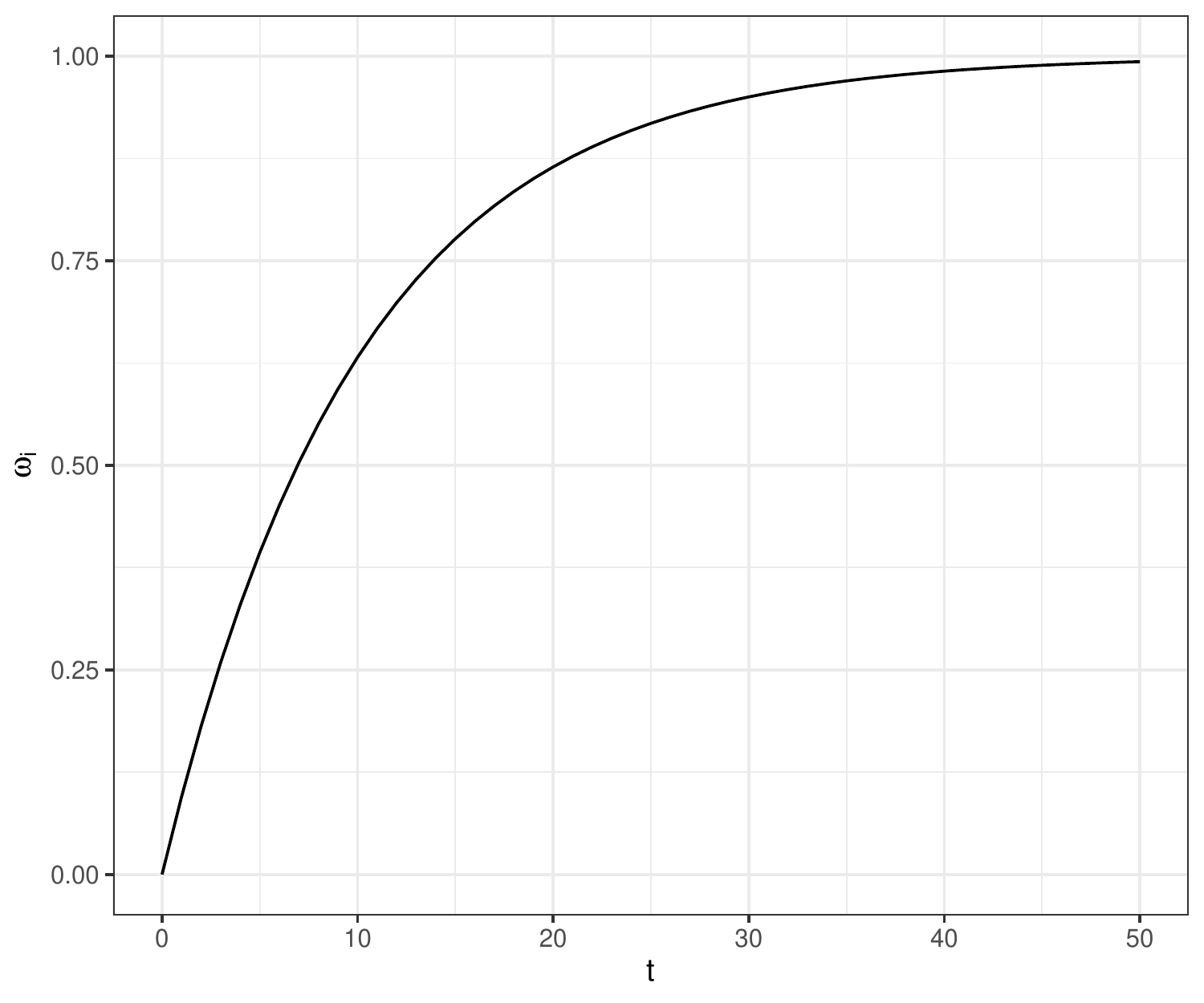}}
\caption{Synthetic example with paths of MEM iterations for a low-density data point with (blue arrows) and without (red arrows) the proposed step size applied (a). Plot of step size as a function of the number of MEM iterations (b).}
\label{fig:step_size_graph}
\end{figure}

Large jumps can occur during the initial iterations of the algorithm for those data points $x_i$'s located in low-density regions. 
In these cases, since most $z_{ik}$'s are very small, the inverse of $\sum_{i=1}^G z_{ik} \Sigmab_k^{-1}$ in \eqref{eq:mem_mstep_opt} will contain large values (in magnitude). 
As a consequence, an initial data point would shift from the region of the attracting mode to the domain of attraction of a different mode, and therefore to converge to a mode different from the attractor of the data point. 
As an example, consider the data point at the bottom-right of Figure~\ref{fig:step_size_graph}a, and the corresponding path of MEM iterations (see the red arrows). In this case, after an initial large jump, the algorithm converges to a mode further from the domain of attraction of the data point.

To avoid these situations, we may compute the update at iteration $t$ as the convex linear combination of the solution at previous step and the proposed value as follows
\begin{equation*}
\x_i^{(t)} = (1-\omega_i)\, \x_i^{(t-1)} + \omega_i\, \x_i^*,
\end{equation*}
where $\omega_i$ is a tuning parameter that controls the step size. 
The definition of $\omega_i$ should consider whether or not a data point lies in a low-density region, and update this value at each iteration. For instance, we could define a function that depends on $\delta_i = \left\vert \sum_{k=1}^G z_{ik} \Sigmab_k^{-1} \right\vert$ for $i=1,\ldots,n$, so when the determinant $\delta_i$ is small, i.e. the corresponding data point lies in a low-density region, the value of $\omega_i$ should be close to zero and the value $\x_i^{(t)}$ should be updated by small steps, whereas for relatively large values of $\delta_i$ the associated weights $\omega_i$ converge to one, so essentially setting $\x_i^{(t)} = \x_i^*$. However, implementing such strategy would require the computation of $\delta_i$ at each iteration for all the data points, and this may result in a significant increase of the execution time. 


For this reason, in practice, we suggest to compute the step size as $\omega_i = 1 - \exp\{-0.1t\}$ (see the function drawn in Figure~\ref{fig:step_size_graph}b). 
The idea is that at earlier iterations, the step size is small and $\x_i^{(t)}$ must be updated by small steps, but as the number of iterations increase the step size converges to one, so the updated value becomes almost equivalent to $\x_i^*$. 
Returning to the example given in Figure~\ref{fig:step_size_graph}a, smaller initial steps (shown as blue arrows) allow to converge to the correct density mode.
Finally, we note that such a strategy inevitably increases the number of steps performed by the MEM algorithm. However, since each step is quite fast to perform, overall the algorithm is not significantly affected. For instance, in the previous example, the MEM iterations for all data points increase from $7$ to $18$, while the execution time from $0.04$ to $0.08$ seconds.

\subsection{Connected-components algorithm for ``tight clusters''}
\label{sec:MEMGMM_concomp}

After the final iteration of the Modal EM algorithm a set of points $\{ \x_i^* \}_{i=1}^n$ are obtained. These represent the modes to which each of the data points converge.
However, in the limit, points that would converge to the same mode may be numerically different from each other by a small amount, whose magnitude depends on the tolerance value used for checking the convergence of the algorithm.
Thus, solutions $\{ \x_i^* \}_{i=1}^n$ form tight clusters around the corresponding modes, widely separated from other tight clusters corresponding to different modes.
The connected-components algorithm described in \citet{Carreira-Perpinan:2016} allows for the merging of those points that ideally would be identical. This can be applied as a post-processing step to obtain the final estimated modes $\{ \hat{\x}_m \}_{m=1}^M$.

\subsection{Denoising low-density modes}
\label{sec:MEMGMM_denoising}

Certain regions of the features space may lack of sufficient data points to obtain reliable density estimates, particularly in high-dimensional features space. As a consequence, modes located in such regions might be spurious and, in these cases, it may be convenient to filter out these modes \citep{Carreira-Perpinan:2000}. 
We consider a simple rule to drop modes associated with regions of relatively low-probability. Following the approach of \citet{Banfield:Raftery:1993}, we postulate the presence of a noise component uniformly distributed over the data region. 
Let $V$ be the hypervolume of the data region, so each log-density value of a mode not exceeding $-\log(V)$ can be considered as a noisy artefact of the density estimation process. Here, the logarithmic scale is used to improve stability and numerical accuracy.

In practice, we need to compute $\log(V)$, and a simple approximation could be obtained by taking the minimum between: (i) the volume of hyperbox containing the observed data; (ii) the volume of the hyperbox obtained from principal component scores; (iii) the volume of ellipsoid hull, i.e. the ellipsoid of minimal volume such that all data points lie inside or on the boundary of the ellipsoid. 
Alternatively, the central $(1-\alpha)100\%$ region of a multivariate Gaussian distribution, i.e. the smallest region such that an observation falls in this region with probability $(1-\alpha)$, can be computed. This region is an ellipsoid in $d$ dimensions, with log-hypervolume equal to 
$$
\log(V) = \log(2) + \frac{d}{2}\log(\pi) - \log(d) - \log\Gamma\left(\frac{d}{2}\right) + \frac{d}{2}\log(\chi^2_{\alpha}(d)) + \frac{1}{2}\log|\Sigmab|,
$$
where $\chi^2_{1-\alpha}(d)$ is the $(1-\alpha)100\%$ quantile of a chi-squared distribution with $d$ degrees of freedom, and $\Gamma()$ the gamma function.
The covariance matrix $\Sigmab$ can be estimated as the marginal covariance matrix using the well-known relationship between the parameters of a multivariate mixture distribution and the marginal parameters \citep[see for instance][Sec. 6.1.1]{FruhwirthSchnatter:2006}, that is
$$
\Sigmab = \sum_{k=1}^G \pi_k \Sigmab_k + \sum_{k=1}^G \pi_k (\mub_k - \mub) (\mub_k - \mub)\T
$$
where $\mub = \sum_{k=1}^G \pi_k\mub_k$ is the vector of marginal means.

Thus, modes whose log-density is smaller than $-\log(V)$ or, equivalently, with density smaller than $\exp(-\log(V)) = 1/V$, can be dropped, and points associated with them are re-assigned to the remaining modes with addtional few steps of the MEM algorithm. 
This is coherent with Hartigan's definition of cluster adopted in the paper, because the groups associated with low-density modes lack the main requirement of being a high density cluster. 
The data example in Section~\ref{sec:bankruptcy} illustrates this approach.

\section{Data analysis examples}

\subsection{Simulated data example}

\begin{figure}[htb]
\centering
\subcaptionbox{}{\includegraphics[width=0.33\textwidth]{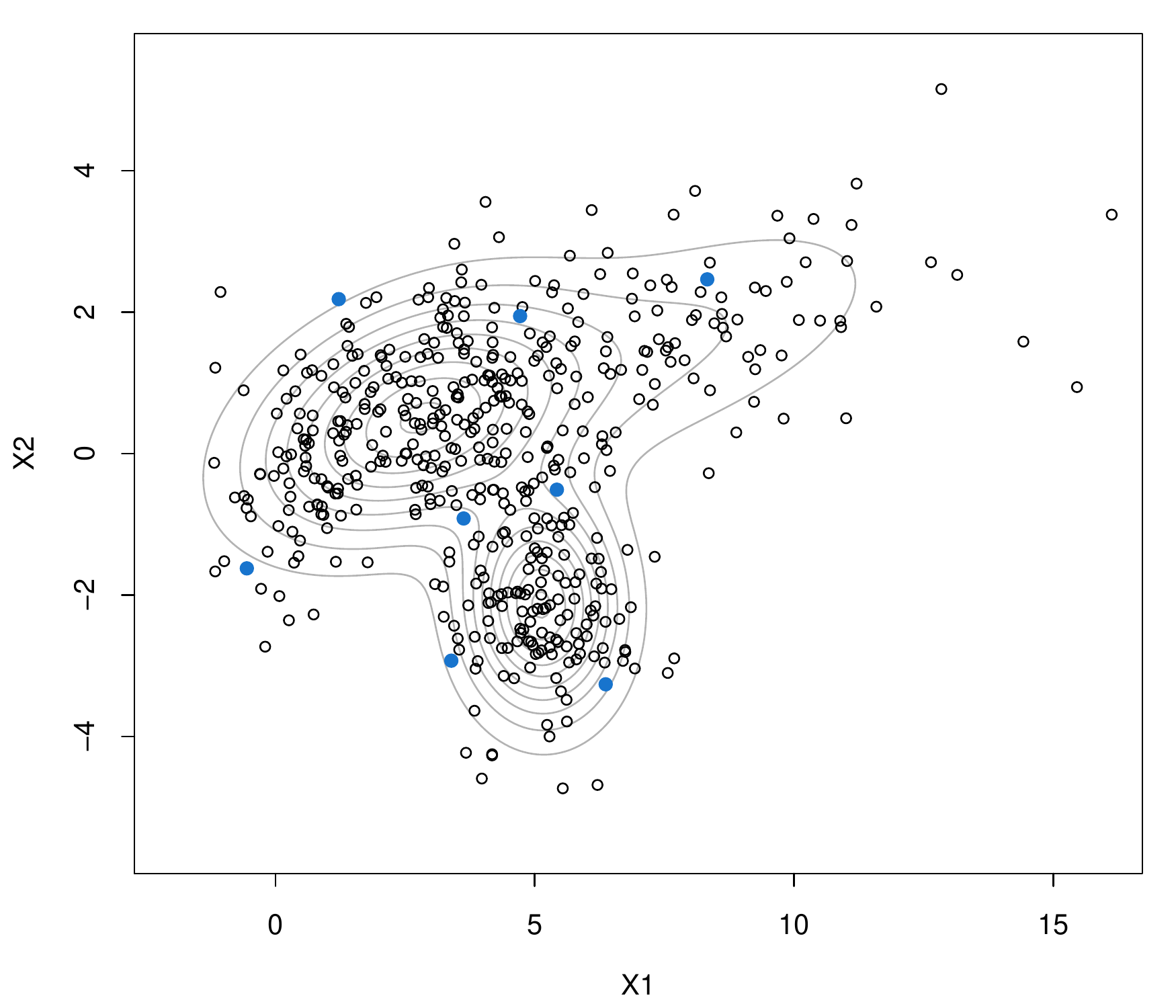}}%
\subcaptionbox{}{\includegraphics[width=0.33\textwidth]{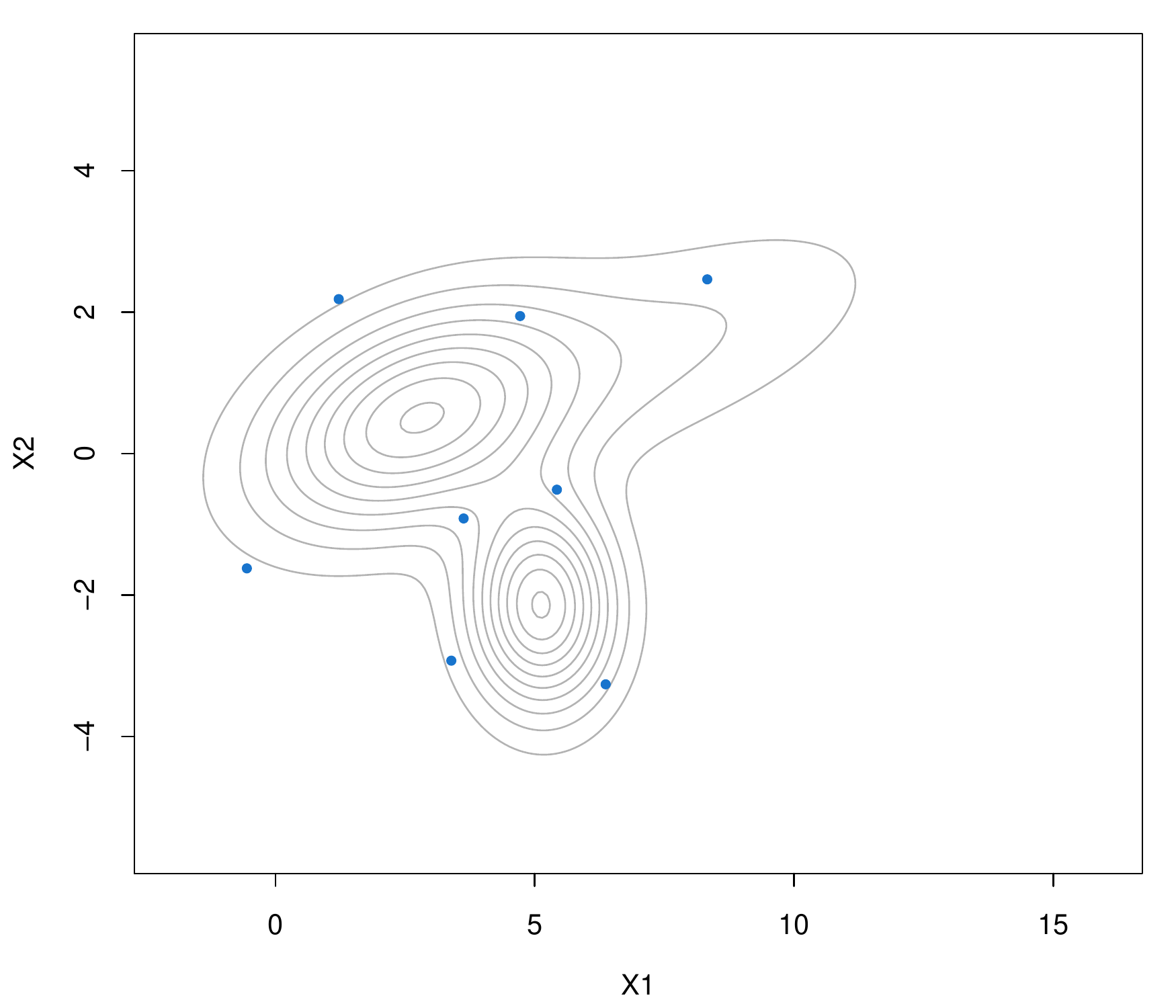}}%
\subcaptionbox{}{\includegraphics[width=0.33\textwidth]{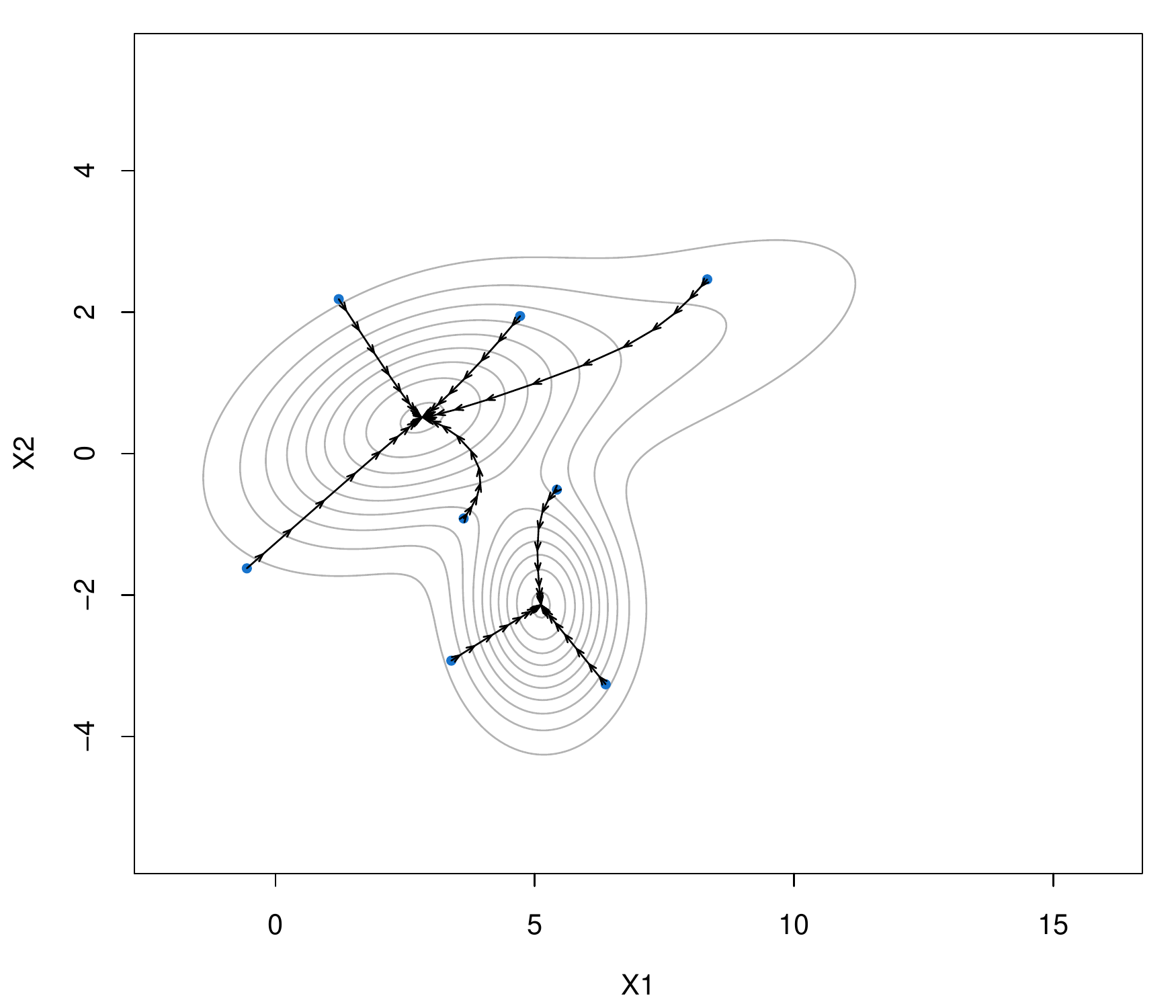}}\\
\subcaptionbox{}{\includegraphics[width=0.33\textwidth]{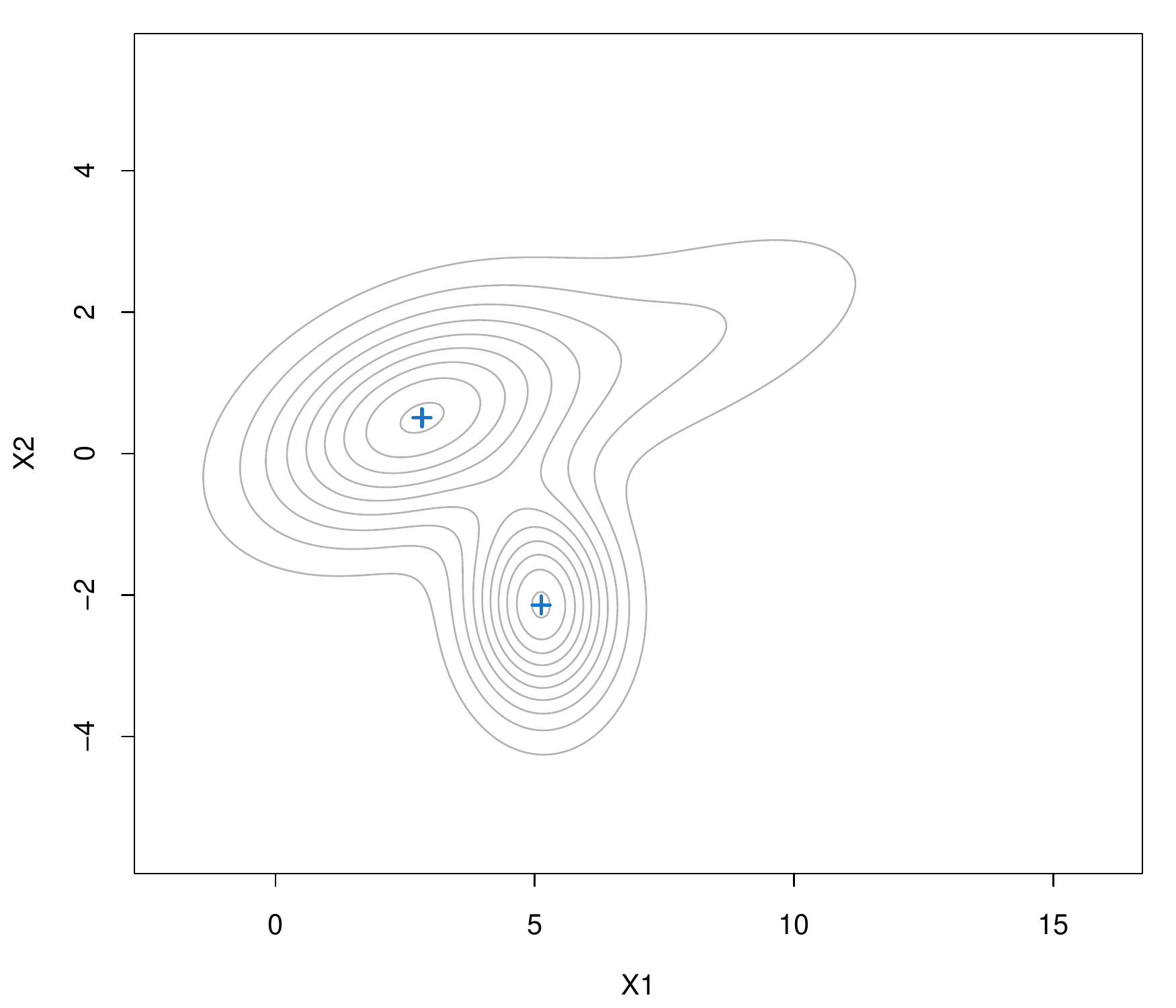}}%
\subcaptionbox{}{\includegraphics[width=0.33\textwidth]{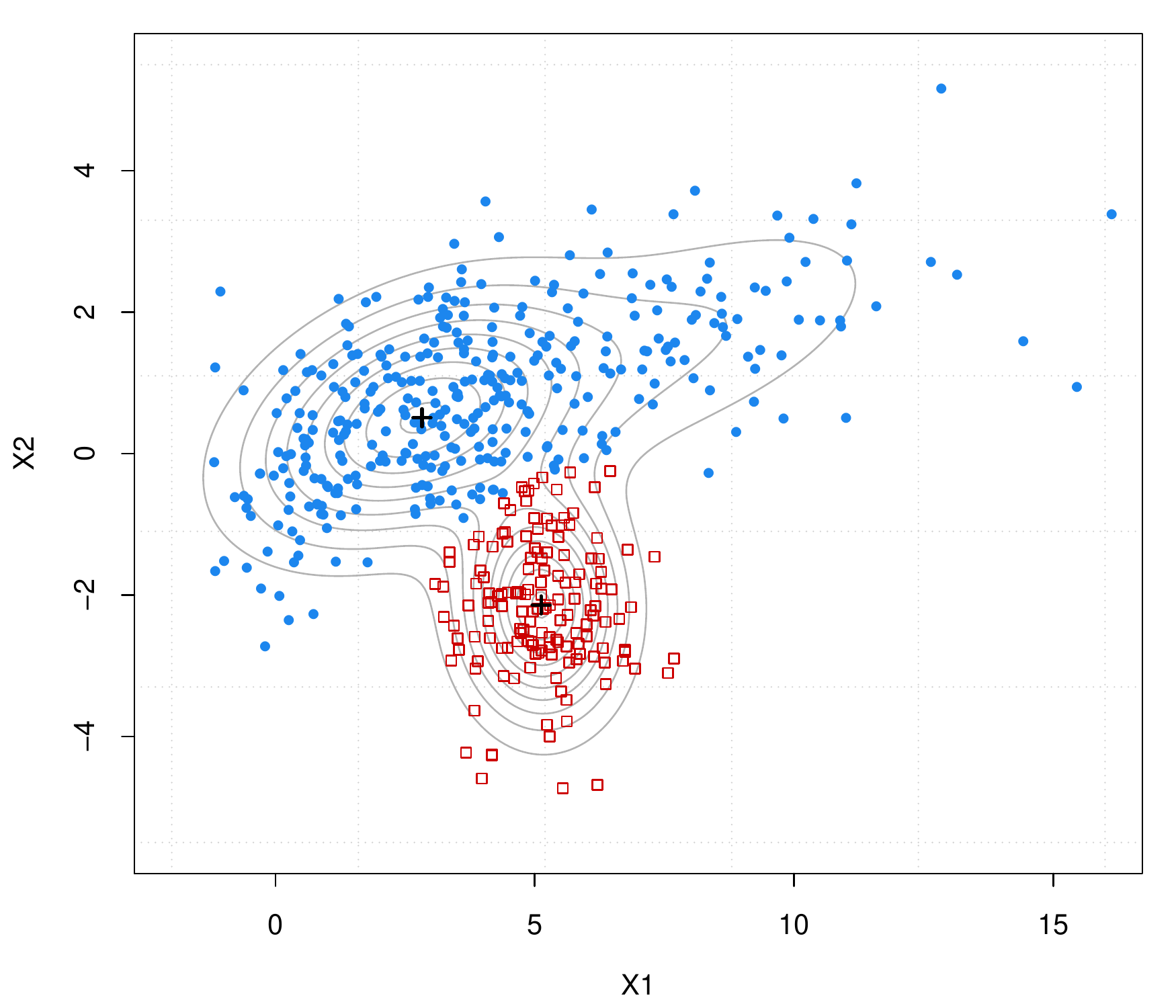}}%
\subcaptionbox{}{\includegraphics[width=0.33\textwidth]{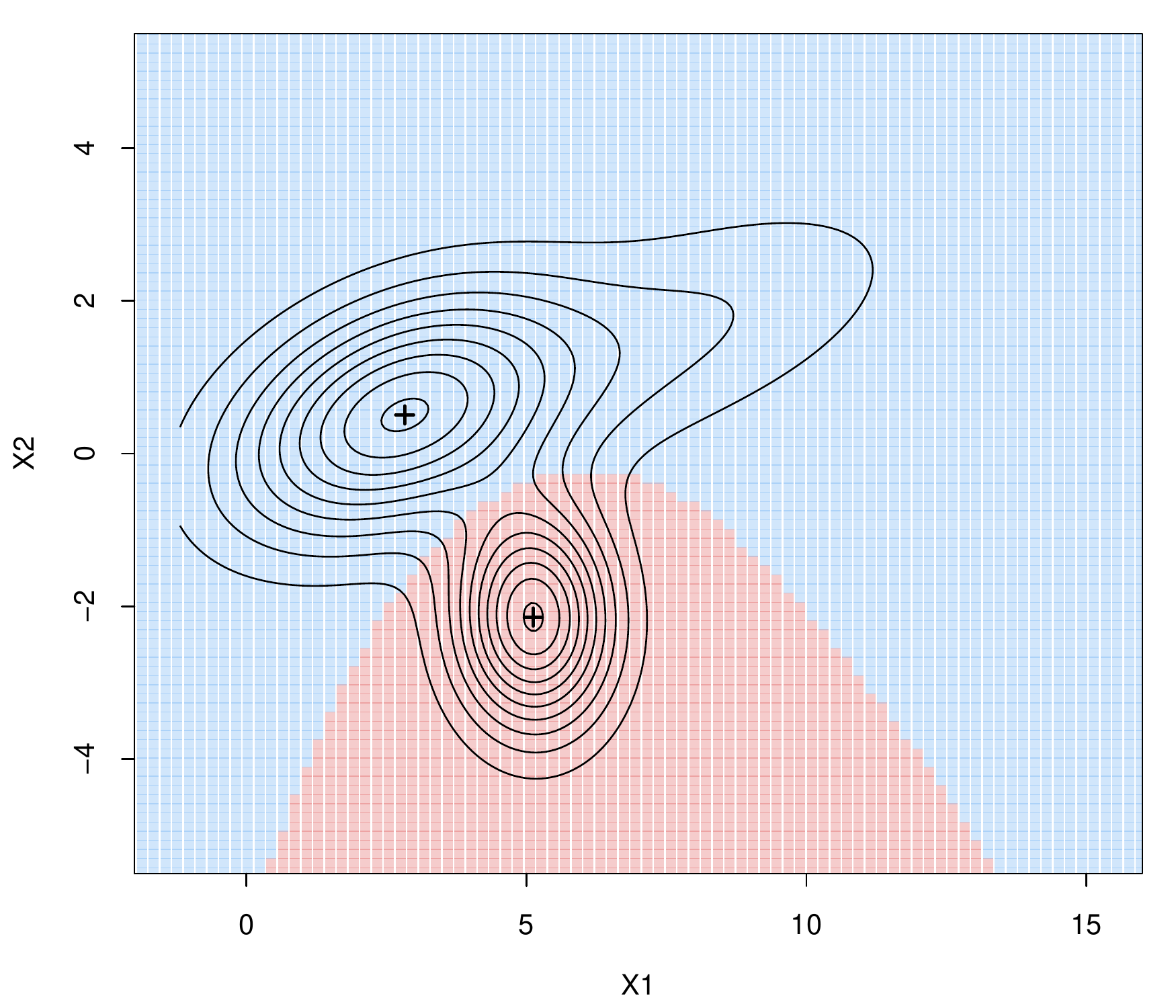}}%
\caption{Plots for the simulated data: (a-b) density contours obtained from the select GMM, with some data points highlighted; (c) paths of MEM algorithm for the selected data points; (d) modes found by the MEM algorithm; (e) modal clustering solution; (f) modal clustering partition of the feature space.}
\label{fig2:motiv_example}
\end{figure}

Recalling the bivariate Gaussian--SkewNormal mixture distribution described in Section~\ref{sec:intro}, Figure~\ref{fig2:motiv_example}a shows the estimated density obtained by the selected Gaussian mixture model, namely model VVE with 3 mixture components selected by BIC.   
To illustrate the procedure, some points are marked as blue filled points, and they are also reported in isolation in Figure~\ref{fig2:motiv_example}b. 
For the selected points, the paths produced by the MEM algorithm described in 
Section~\ref{sec:MEMGMM} are shown as arrows in Figure~\ref{fig2:motiv_example}c. At each step of the algorithm, points move up-hill toward the density modes. The estimated modes are shown in Figure~\ref{fig2:motiv_example}d. 
Figure~\ref{fig2:motiv_example}e shows the modal clustering for all the data points obtained according to the mode to which they converge.
The MEM algorithm required 21 iterations and $0.23$ seconds to run on an iMac with 4 cores i5 Intel CPU running at 2.8 GHz and with 16GB of RAM. 
Finally, Figure~\ref{fig2:motiv_example}f illustrates the partition of the feature space that defines the ``domains of attraction'' of the estimated density modes.

\subsection{Mass cytometry data}
\label{sec:masscytometry}

\begin{figure}[htb]
\centering
\subcaptionbox{}{\includegraphics[width=0.4\textwidth]{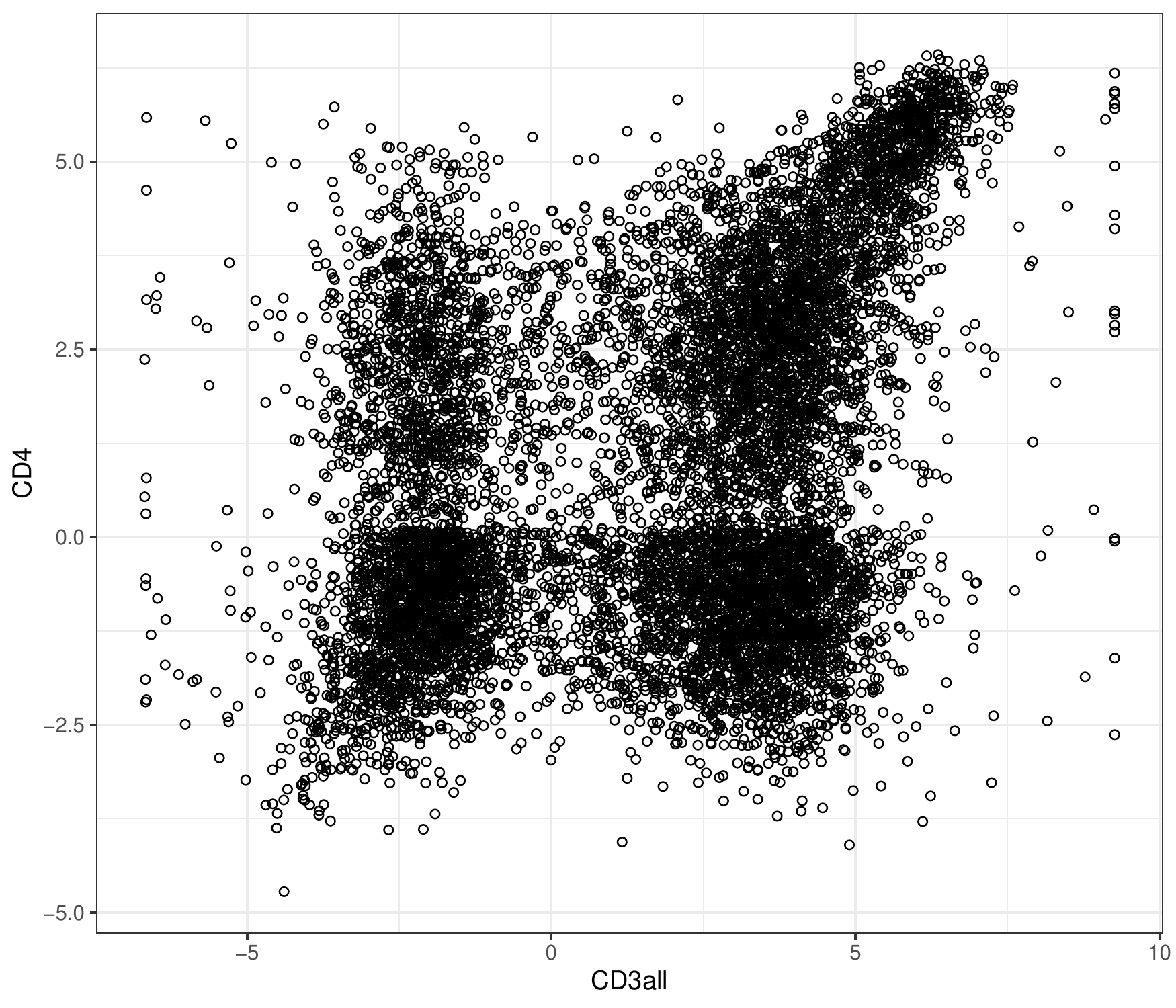}}
\subcaptionbox{}{\includegraphics[width=0.4\textwidth]{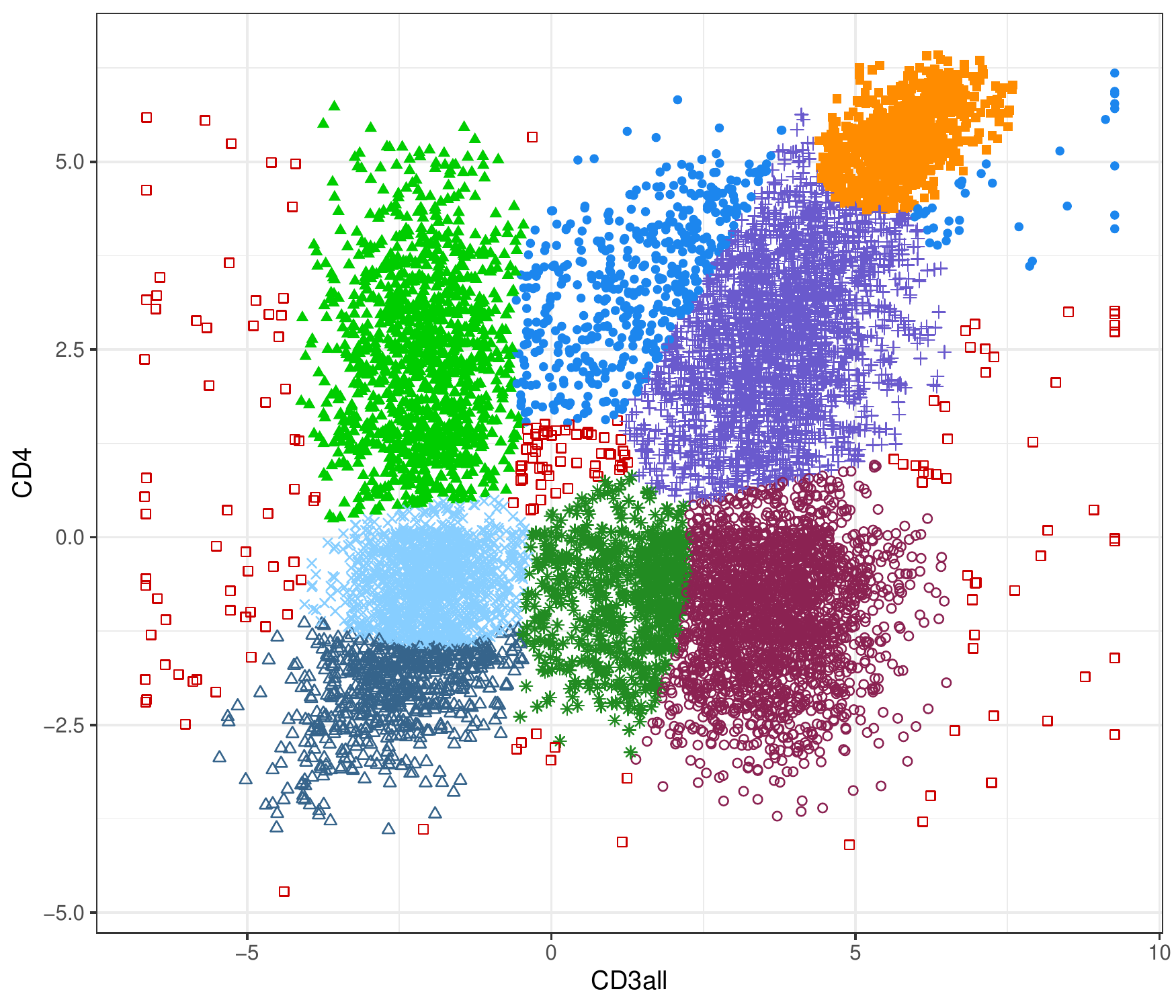}} \\
\subcaptionbox{}{\includegraphics[width=0.4\textwidth]{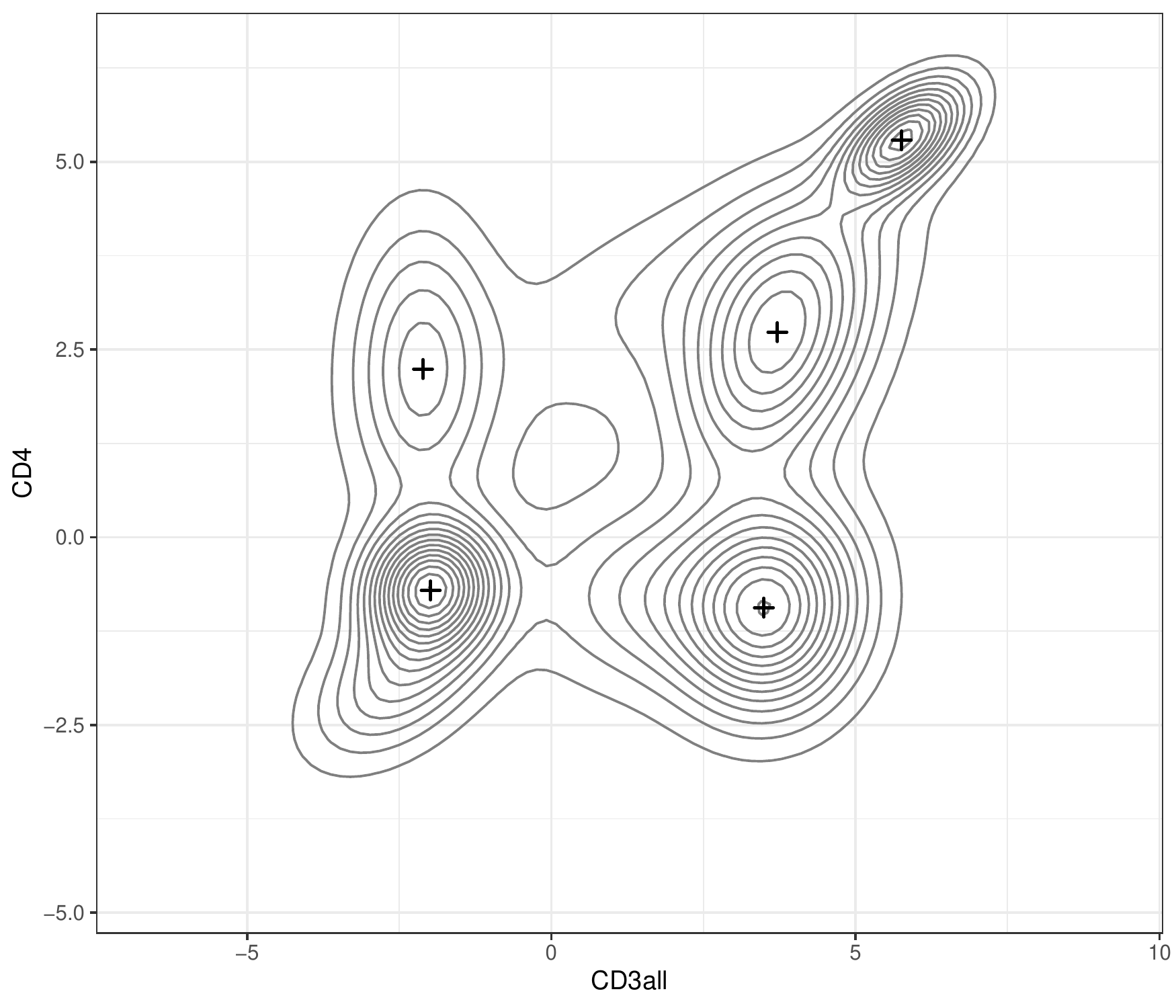}} 
\subcaptionbox{}{\includegraphics[width=0.4\textwidth]{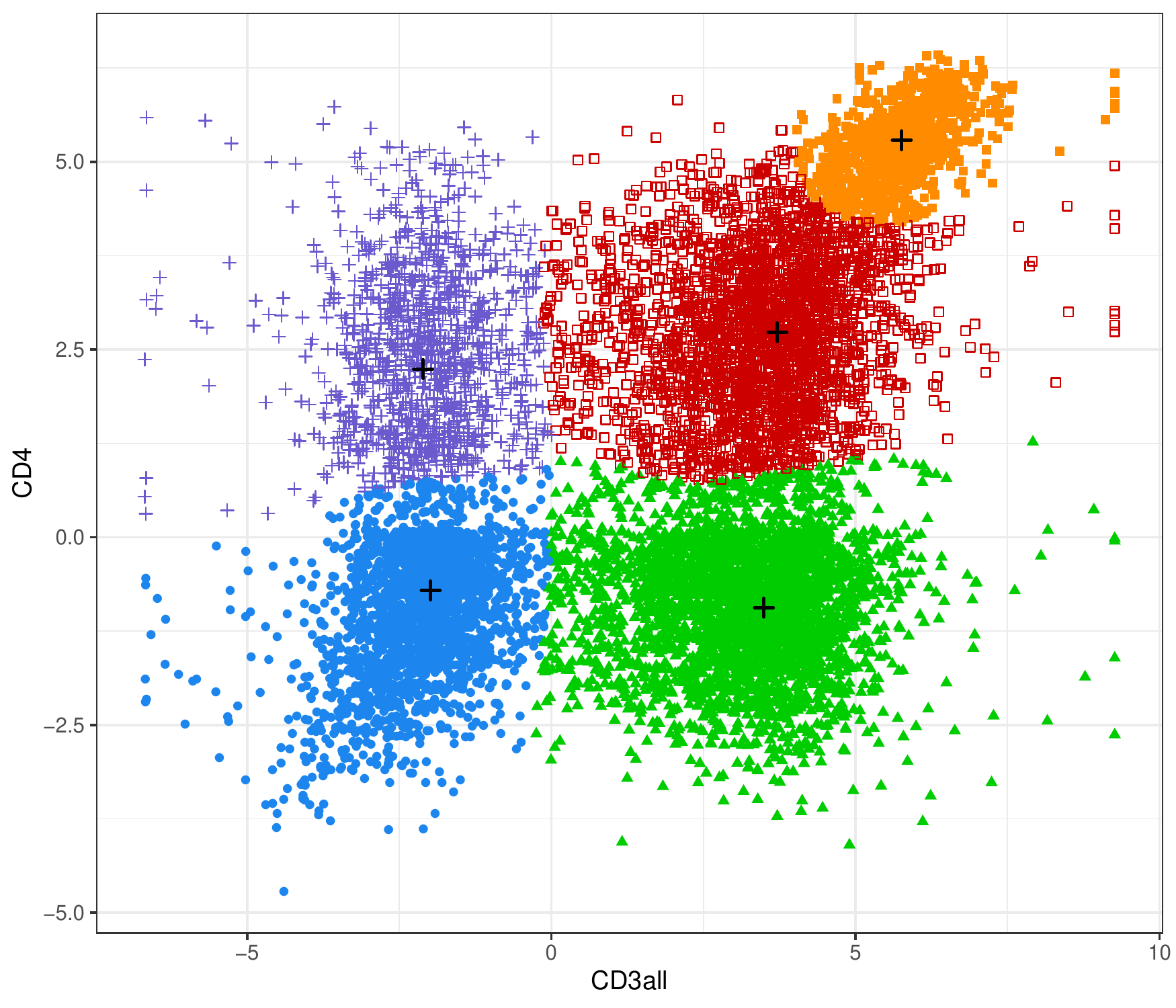}} \\
\subcaptionbox{}{\includegraphics[width=0.4\textwidth]{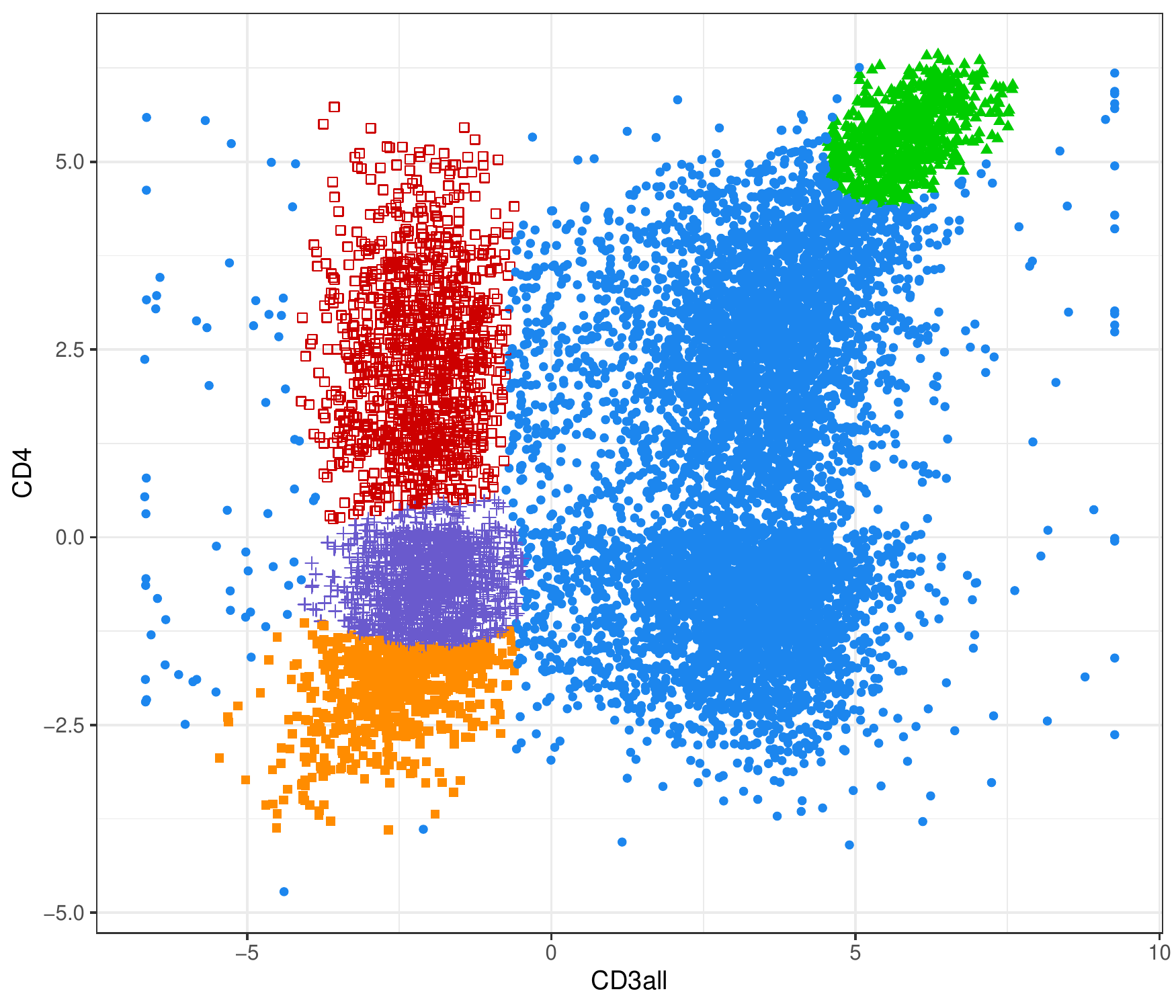}} 
\subcaptionbox{}{\includegraphics[width=0.4\textwidth]{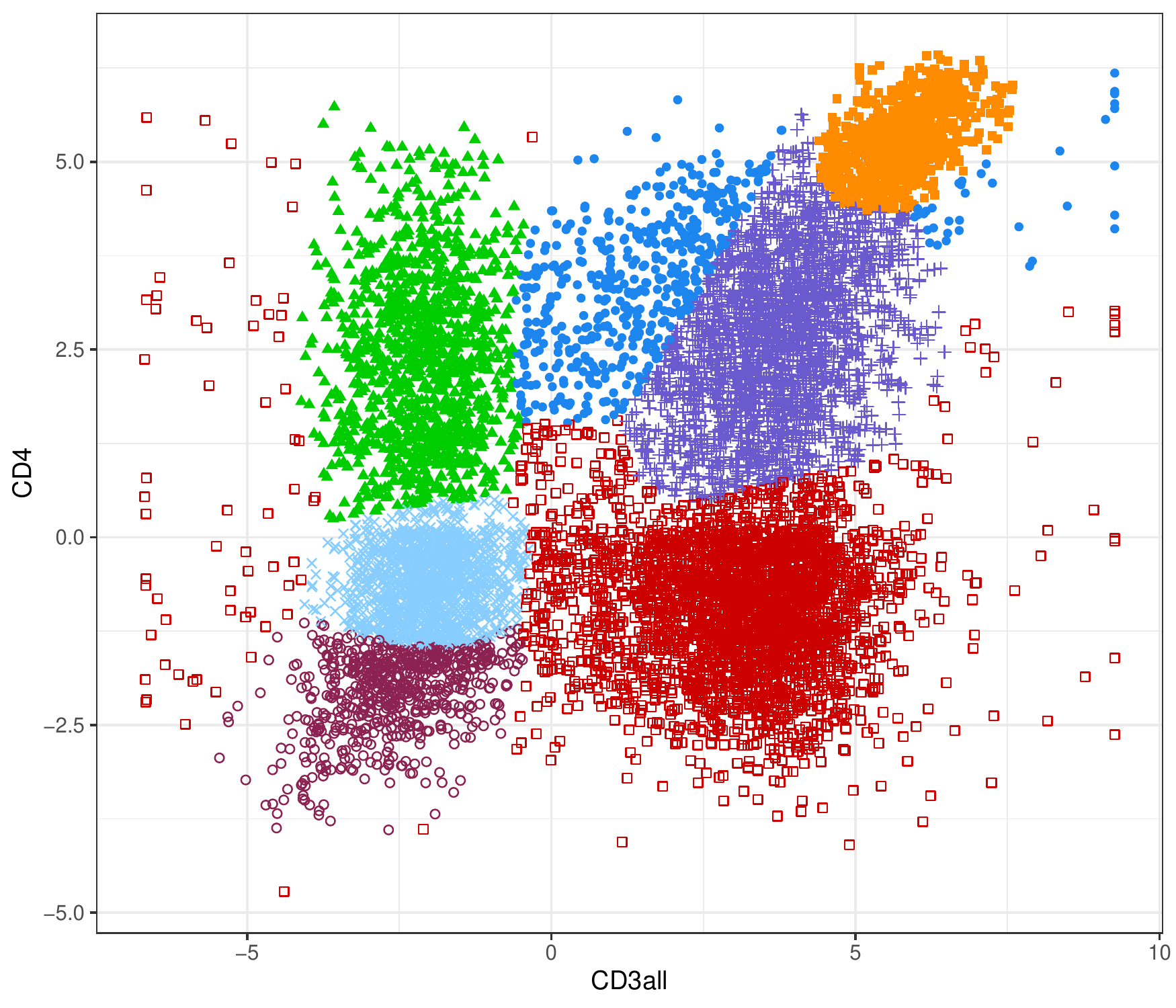}} 
\caption{Mass cytometry experiment data: (a) plot of CD4 and CD3all markers; (b) clustering from the unrestricted GMM with 9 components; (c) contour plot of density estimate and modes (\textbf{+}) estimated by the MEM algorithm; (d) modal clustering classification of cells; (e) clustering obtained by merging mixture components using the \citet{Baudry:etal:2010} approach; (f) clustering obtained by merging mixture components using the unimodal ridgeline approach of \citet{Hennig:2010}.}
\label{fig:masscytometry}
\end{figure}

Mass cytometry is a recent technology that couples flow cytometry with mass spectrometry. It allows to simultaneously measure several features of a cell. 
The biological question of interest is the identification of subpopulations of cells. We consider two protein-markers, CD4 and CD3all, from a mass cytometry experiment \citep{Bendall:etal:2011} to find latent classes in single-cell measurements. Data are preprocessed using the hyperbolic arcsin transformation \citep{Holmes:Huber:2018}, i.e. $\text{asinh}(x) = \log(x + \sqrt{x^2+1})$. 
A random sample of 10,000 cells (out of 91,392) is shown in Figure~\ref{fig:masscytometry}a. The clustering obtained with the "best" GMM selected by BIC -- namely, VVV in the \code{mclust} nomenclature \citep[Table 3]{Scrucca:etal:2016} with 9 components -- is shown in Figure~\ref{fig:masscytometry}b. 
The corresponding density estimate and the modes estimated via the MEM algorithm are reported in Figure~\ref{fig:masscytometry}c, while Figure~\ref{fig:masscytometry}d shows the corresponding modal clustering partition. There appears to be five clusters, one for each combination of high/low values of the CD4 and CD3all markers, and an additional cluster formed by the highest values of both the CD4 and CD3all markers.
This result can be contrasted with those obtained using approaches based on merging mixture components. Figure~\ref{fig:masscytometry}e shows the partition derived from the entropy-based approach of \citet{Baudry:etal:2010},
while Figure~\ref{fig:masscytometry}f shows the clusters obtained using the unimodal ridgeline approach of \citet{Hennig:2010}. Both partitions are clearly different from the one obtained using the MEM algorithm, with the latter producing more compact and easily interpretable clusters.

Finally, we note that with 10 thousands observations the MEM algorithm required 24 iterations and $5.4$ seconds to run on an iMac with 4 cores i5 Intel CPU running at 2.8 GHz and with 16GB of RAM.

\subsection{Bankruptcy dataset}
\label{sec:bankruptcy}

\begin{figure}[htb]
\centering
\subcaptionbox{}{\includegraphics[width=0.4\textwidth]{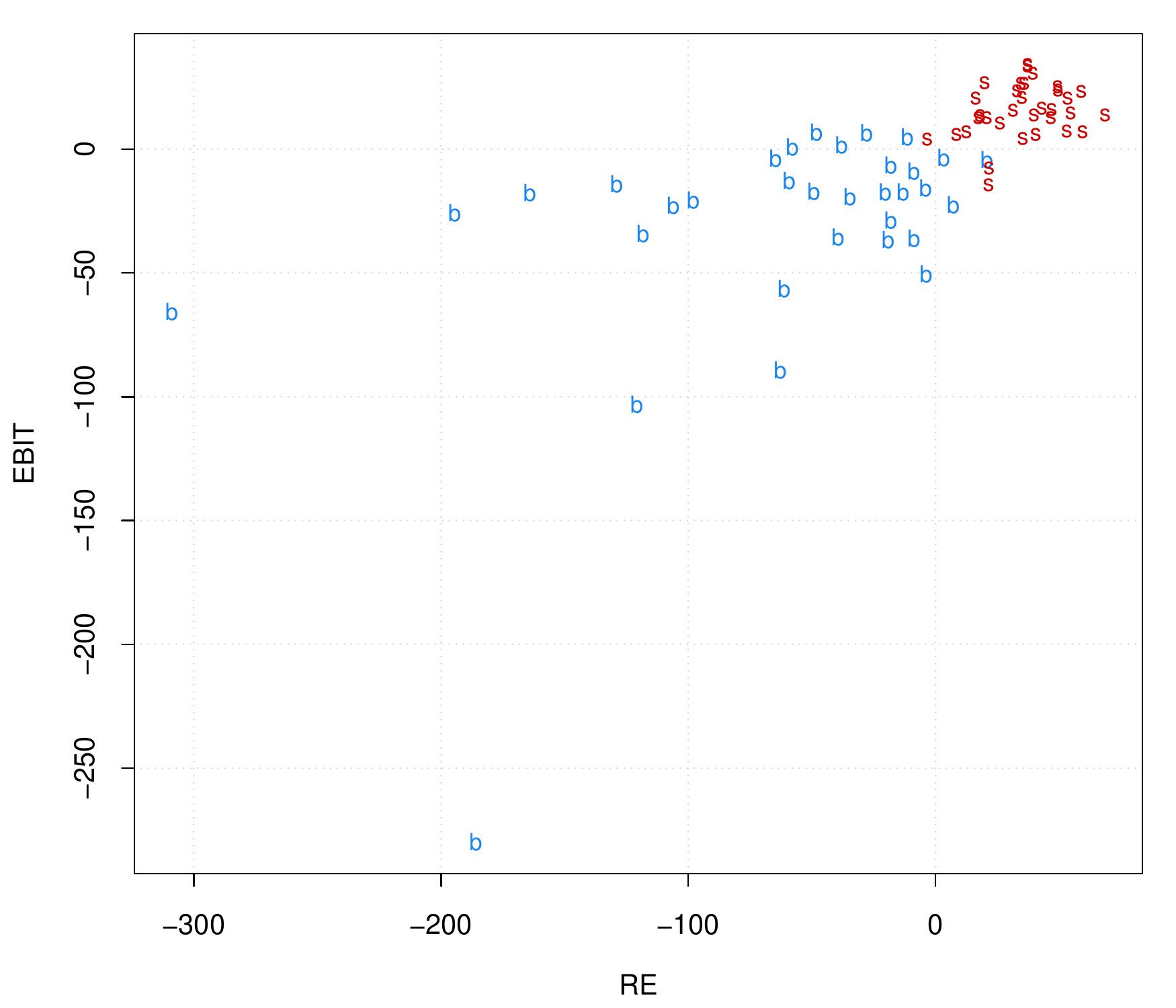}}
\subcaptionbox{}{\includegraphics[width=0.4\textwidth]{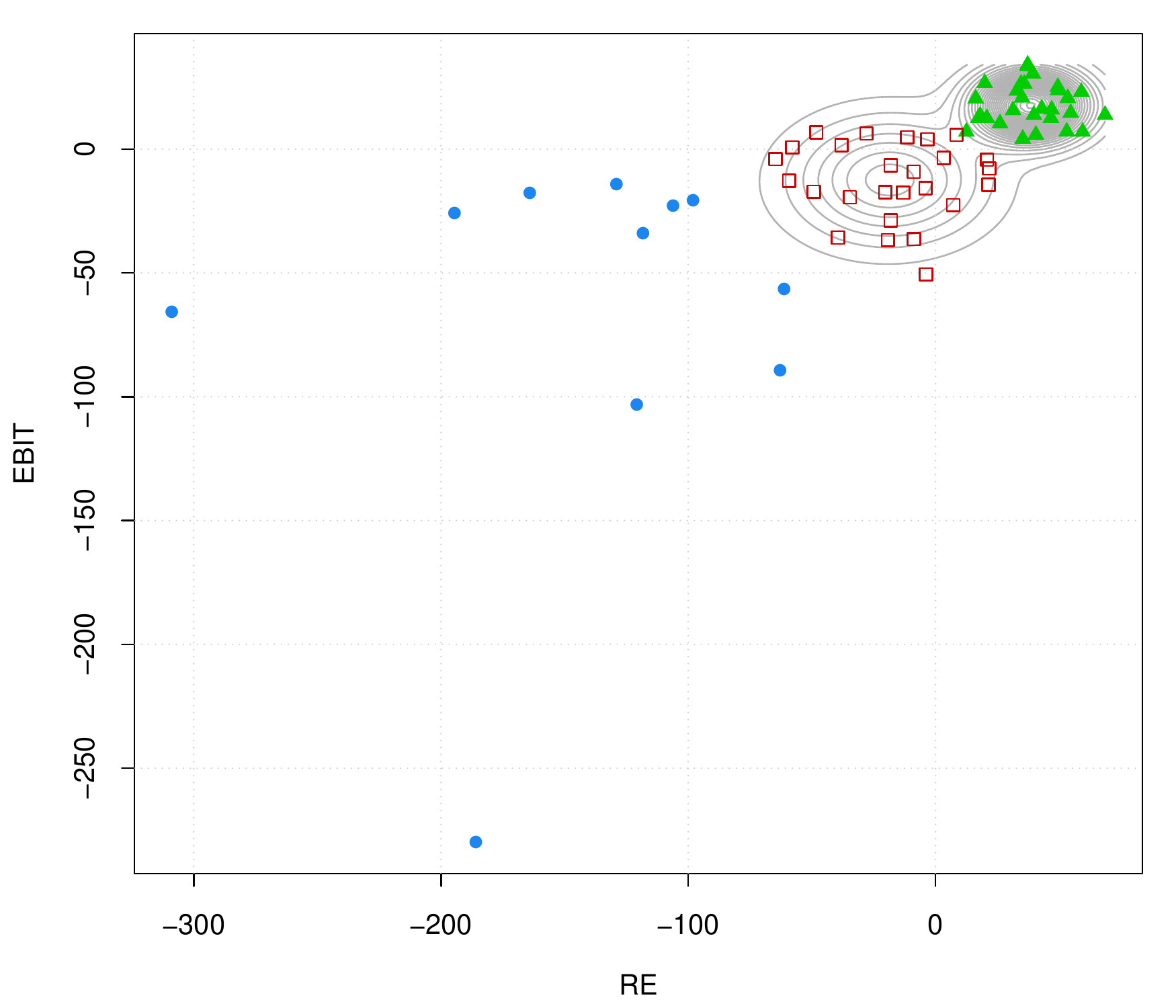}}\\
\subcaptionbox{}{\includegraphics[width=0.4\textwidth]{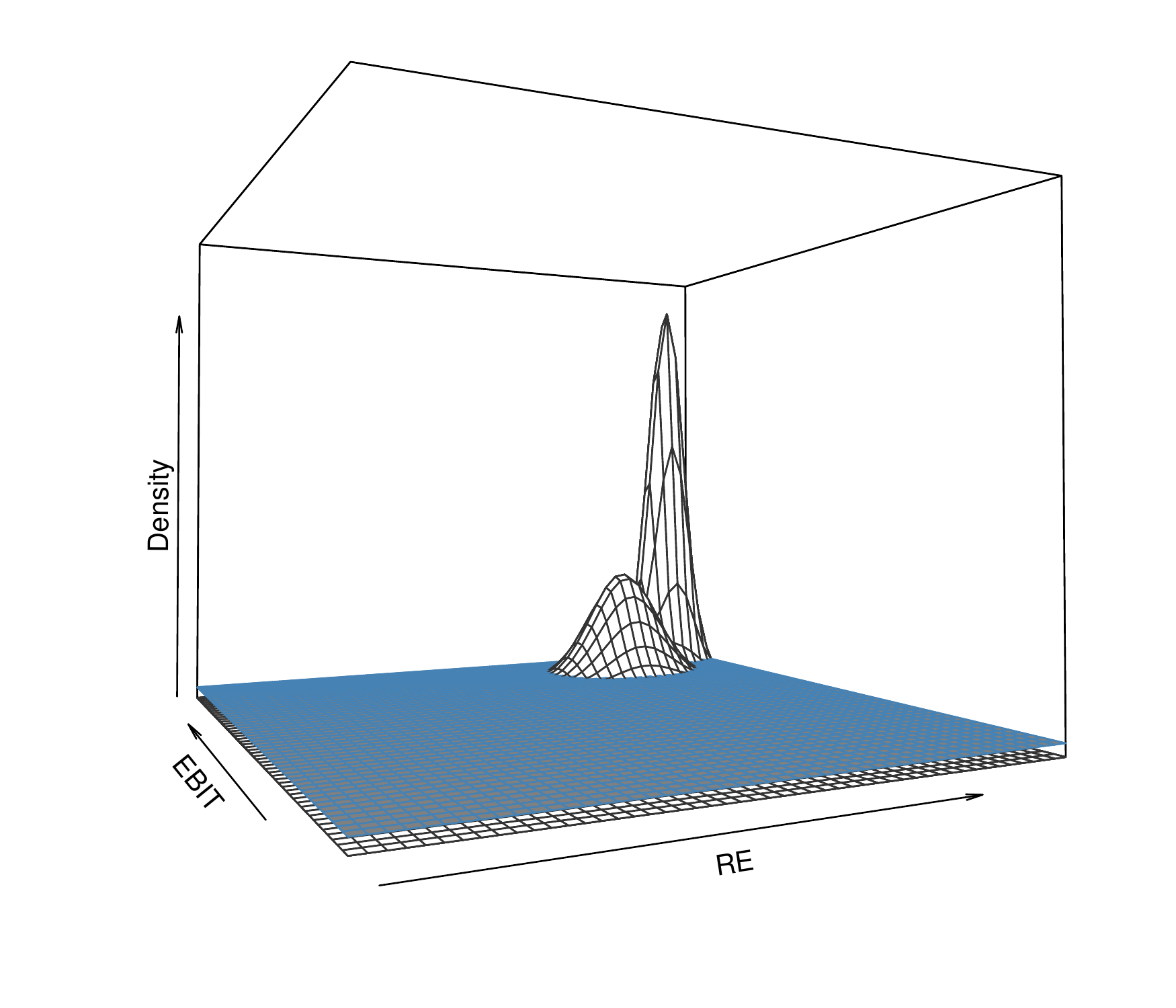}}
\subcaptionbox{}{\includegraphics[width=0.4\textwidth]{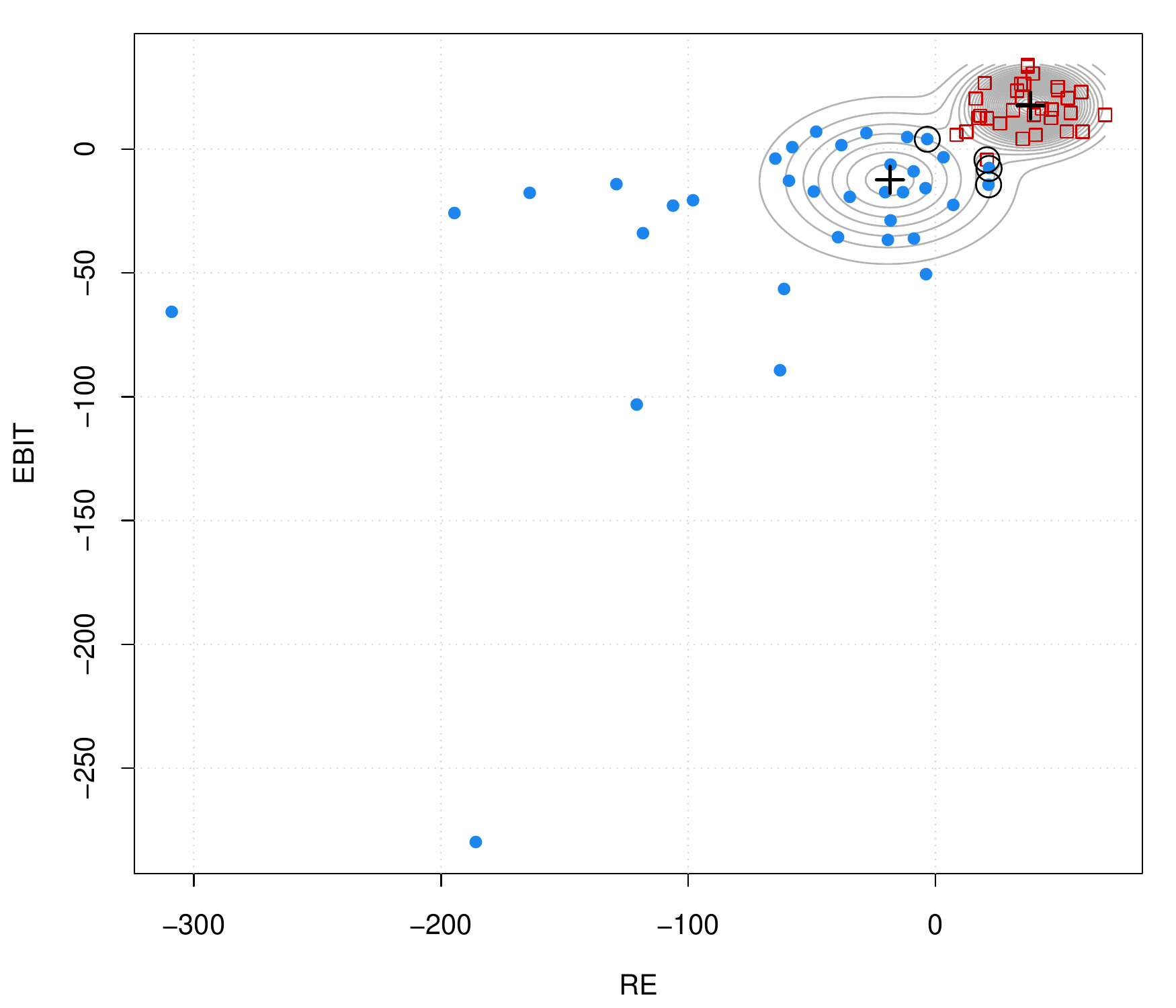}}
\caption{Altman's bankruptcy dataset: (a) financial ratios of companies with points marked as solvent or bankruptcy; (b) contour plot of density estimate and classification obtained by the best fitting GMM as selected by BIC; (c) perspective plot of density estimate and a plane drawn at the uniform density level for denoising; (d) modes and classification obtained from applying the MEM algorithm, with circled points corresponding to misclassified observations.}
\label{fig:bankruptcy}
\end{figure}

\citet{Altman:1968} presented a study on financial ratios to predict corporate bankruptcy. The dataset provides the ratio of retained earnings (RE) to total assets, and the ratio of earnings before interests and taxes (EBIT) to total assets, for a sample of 66 manufacturing US firms, of which 33 had filled for  bankruptcy in the following two years. Data are shown in Figure~\ref{fig:bankruptcy}a.

The best GMM selected by BIC is the model VEI (diagonal, varying volume and equal shape) with three components. The contour plot of the corresponding density estimate is shown in Figure~\ref{fig:bankruptcy}b, with points marked according to the implied maximum a posteriori (MAP) classification. There appears to be two prominent clusters, consisting mainly of solvent and bankrupt companies, but also a spread out group of firms with very low values for either financial ratios.
Figure~\ref{fig:bankruptcy}c shows the density estimate via a 3D perspective plot. Following the approach described in Section~\ref{sec:MEMGMM_denoising}, a plane is included at the uniform density level corresponding to $\exp(-11.17492) = 1/71319.39 = 1.402\times 10^{-5}$, where $V = 71319.39$ is the hypervolume of the $99\%$ central region. 
As it can be seen, only two bumps of densities emerge, namely those corresponding to the main groups in the data. A very low density mode is also present in the region corresponding to small financial ratios; specifically, the density is equal to $4.661\times 10^{-6}$, a value approximately equal to one third the density threshold computed above. For this reason, the low density mode is filtered out by the denoising procedure.

The modes estimated by the MEM algorithm are shown in Figure~\ref{fig:bankruptcy}d, with data points marked according to the clusters assigned by the modal clustering procedure. Only 4 companies are misclassified, as indicated by the circled data points. This result can be compared with those reported by \citet[Table 1]{Lo:Gottardo:2012}, where the best model (a mixture of $t$ distributions on Box-Cox transformed data) misclassified 10 observations.

\clearpage

\section{Conclusions} 
\label{sec:conclusions}

This paper addresses the problem of computing the modes of a density estimated by fitting a Gaussian mixture model. The proposed approach is based on the Modal EM algorithm, an iterative procedure aimed at identifying the local maxima of a density function. By exploiting specific characteristics of the underlying Gaussian mixture model, we extend the Modal EM algorithm to deal with any parsimonious component-covariance matrix decomposition. Furthermore, we discuss a fast implementation of the algorithm that allows to perform the M-step simultaneously for all data points.
Once the modes of the underlying density are estimated, a modal clustering partition can be obtained by associating each observation to the pertaining mode.

The Modal EM algorithm discussed, as any other mode-seeking procedure, relies on the quality of the underlying density estimate. Clearly, if the parameters of the mixture model are not well estimated some issues could arise. 
However, provided that the general form of the density estimate is not overly biased, the proposed method should not be significantly affected. 

The proposed approach seems to be very promising and, in principle, it could be extended to mixtures of non-Gaussian distributions (e.g. $t$, skew-Normal, skew-$t$, shifted asymmetric Laplace, \ldots). However, it is necessary to investigate the potential benefits obtained by adopting more complex probability models.
Recently, an adaptation of the proposed algorithm has been used for clustering from an ensemble of Gaussian mixtures \citep{Casa:Scrucca:Menardi:2019}.

Another area of future research involves the use of the MEM algorithm in high-dimensional data settings. In this regard, we plan to study the effectiveness of the MEM algorithm applied to the subspace estimated by the GMM-based projection pursuit method proposed by \citet{Scrucca:Serafini:2019}.

\bibliographystyle{apalike}
\bibliography{MEMGMM_arxiv}

\end{document}